\documentclass[a4paper,aps,prd,preprint,showkeys,nofootinbib,superscriptaddress]{revtex4-1}
\usepackage{color}
\usepackage{amsmath}
\usepackage{amssymb}
\usepackage{graphicx}
\usepackage[colorlinks=true, linkcolor=blue, citecolor=green]{hyperref}
\graphicspath{{figs/}}
\usepackage[caption=false]{subfig}
\begin{document}
\title{Holographic superfluid solitons with backreaction}
\author{Zhongshan Xu}
\email{xuzhongshan16@mails.ucas.edu.cn}
\affiliation{School of Physical Sciences, University of Chinese Academy of Sciences, Beijing 100049, China}
\author{Yiqiang Du}
\email{ydu@physik.uni-wuerzburg.de}
\affiliation{Institute for Theoretical Physics and Astrophysics and W\"urzburg-Dresden Cluster of Excellence ct.qmat, Am Hubland 97074 W\"{u}rzburg, Germany}
\affiliation{Department of Physics, Hanyang University, 222 Wangshimni-ro, Sungdong-gu, Seoul, 04763, South Korea}
\author{Johanna Erdmenger}
\email{erdmenger@physik.uni-wuerzburg.de}
\affiliation{Institute for Theoretical Physics and Astrophysics and W\"urzburg-Dresden Cluster of Excellence ct.qmat, Am Hubland 97074 W\"{u}rzburg, Germany}
\author{Ren\'{e} Meyer}
\email{rene.meyer@physik.uni-wuerzburg.de}
\affiliation{Institute for Theoretical Physics and Astrophysics and W\"urzburg-Dresden Cluster of Excellence ct.qmat, Am Hubland 97074 W\"{u}rzburg, Germany}
\author{Yu Tian}
\email{ytian@ucas.ac.cn}
\affiliation{School of Physical Sciences, University of Chinese Academy of Sciences, Beijing 100049, China}
\affiliation{Center for Theoretical Physics, Massachusetts Institute of Technology,
Cambridge, Massachusetts 02139, USA}
\affiliation{Institute of Theoretical Physics, Chinese Academy of Sciences, Beijing
100190, China}
\author{Zhuo-Yu Xian}
\email{xianzy@itp.ac.cn}
\affiliation{Institute of Theoretical Physics, Chinese Academy of Sciences, Beijing
100190, China}
\begin{abstract}
Solitons are important nonperturbative excitations  in superfluids. For holographic superfluids, we numerically construct ``dark'' solitons that have the symmetry-restored phase at their core. A central point is that we include the gravitational backreaction of the matter fields, which becomes important at low temperatures. We study in detail the properties of these solitons under variation of the backreaction strength via tuning the gravitational constant. In particular, the depletion fraction of the particle number density at the core of the solitons is carefully investigated. In agreement with the probe-limit analysis, the depletion fraction shows the same qualitative behavior as in Bogoliubov-de Gennes theory, even if the backreaction is included. We find that the depletion decreases with increasing backreaction strength. Moreover, the inclusion of backreaction enables us to obtain the effective energy density of solitons within holography, which together with an evaluation of the surface tension leads to a simple physical explanation for the snake instability of dark solitons. 
\end{abstract}
%\keywords{Soliton, Holography, Back-reaction}
\maketitle
\tableofcontents

\section{Introduction}

Gauge/gravity duality \cite{Maldacena1998,Witten1998,Gubser1998} is a powerful tool to describe strongly coupled and correlated systems. Many problems associated with strongly interacting condensed matter physics are tractable in this setup \cite{Johanna2015}. One of these problems is unconventional  superfluidity \cite{Herzog2009,Hartnoll2008}. 

\begin{figure}
\centering
\includegraphics{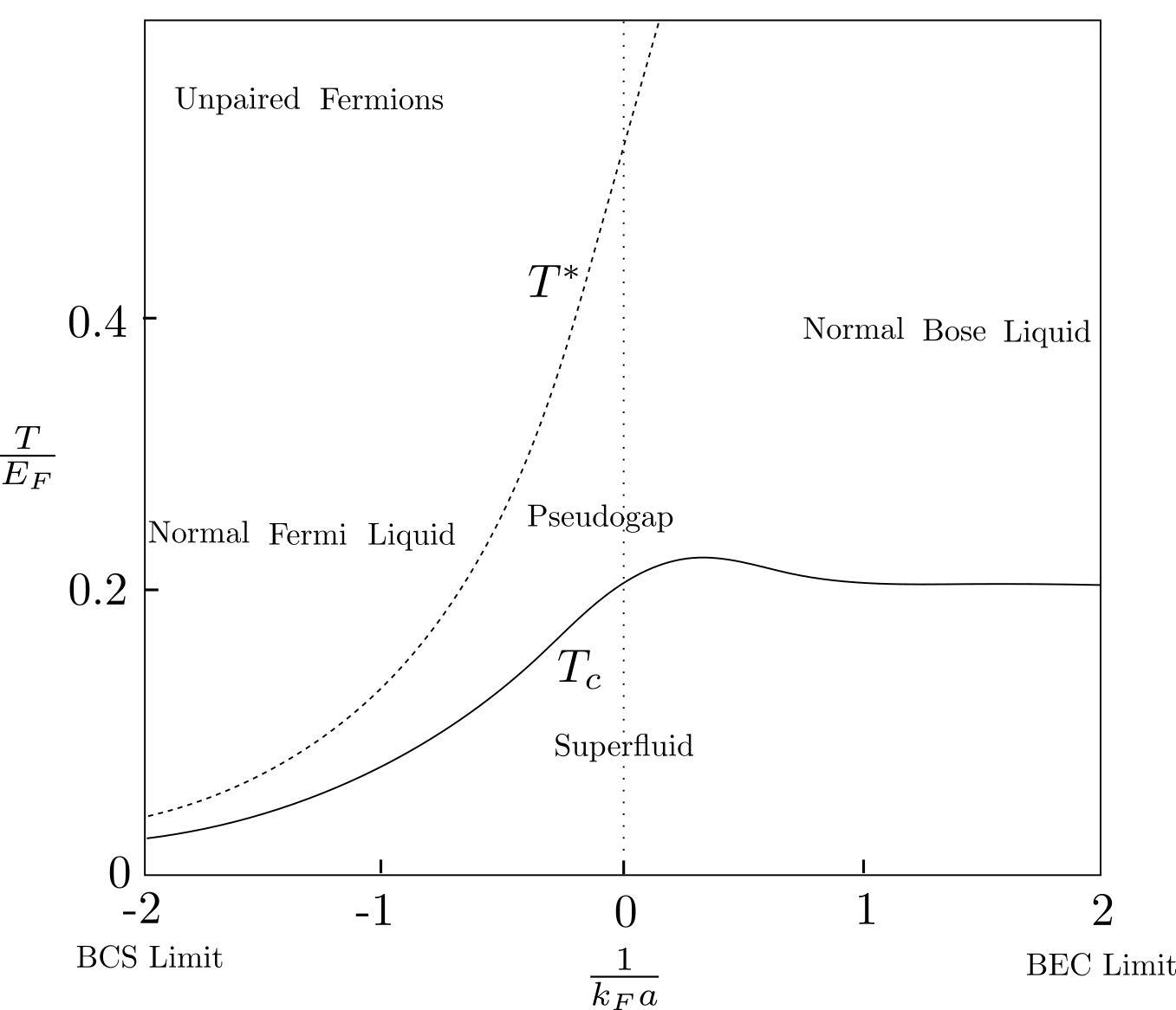}
\caption{\label{fig:phase diagram}Qualitative phase diagram reproduced from
Ref.~\cite{Randeria2014} of the BCS-BEC crossover as a function of temperature $T/E_{F}$ and coupling $1/k_{F}a$, where $k_{F}$ is the Fermi momentum and $a$ is the scattering length. $T^{\ast}$ is the pairing onset temperature, while $T_{c}$ is the critical temperature for the superfluidity.}
\end{figure}

Superfluidity is a collective quantum phenomenon occurring in both bosonic and fermionic systems at low temperatures. In particular, fermionic systems can interpolate in a smooth way between the formation and condensation of loosely bound Cooper pairs (BCS superconductivity), and Bose-Einstein condensation (BEC) of preformed bosonic molecules. This is known as the BCS-BEC crossover (see Ref.~\cite{Randeria2014} for a review), which has been realized by cooling fermionic gases to ultra-low temperatures and tuning the interactions between the fermions with a controllable external magnetic field in the laboratory. The qualitative essence of this crossover can be understood from the phase diagram depicted in Fig.~\ref{fig:phase diagram}: the horizontal axis interpolates between the BCS regime of a weakly attractive interaction between fermions and the BEC regime of very strong attraction. Above the pairing onset temperature $T^{\ast}$, the system is a normal Fermi liquid consisting of the unpaired fermions on the BCS side. On the BEC side, the the strong attractive interaction binds the fermions together to form bosonic molecules, which at large enough temperatures form a Bose liquid. As the temperature on the BCS side decreases, loosely bound Cooper pairs of fermions start to form at $T<T^{\ast}$  and  condense below a critical temperature $T_c$. On the BEC side, Bose-Einstein condensation of the bosonic molecules occurs for $T<T_{c}$. Between these two regimes and at $T<T_{c}$, there exists a strongly coupled regime of unconventional  superfluidity around the point of infinite scattering length $\frac{1}{k_{F}a} =  0$, which is known as the unitary Fermi gas.

Superfluids are quantum fluids which can sustain nonperturbative excitation such as  solitons and vortices \cite{Esko2010,Esko2010a}. One kind of soliton is the so-called dark soliton, which interpolates between the two phases of the system\footnote{Dark solitons include gray solitons with definite velocity and black solitons that are static \cite{Frantzeskakis2010}. We focus on black solitons in this work.}. The core of dark solitons is in the normal state where the condensate vanishes, while asymptotically far away from the core, the soliton is in the superfluid state. The core properties can be used to provide insights into the microscopic structure of the superfluid \cite{Pitaevskii2007}. It was found that, at zero temperature, solitons display distinctively different core properties in the BEC and BCS regimes. For the BEC superfluid, solitons have vanishing core particle number density, because bosonic molecules fully condense below the critical temperature, and there are no non-condensed states available. But in the core of the BCS dark soliton, there are still unpaired fermionic atoms or loosely bound Cooper pairs in excited states, which lead to a non-vanishing core particle number density. At finite temperature, even in the BEC superfluid there will be a small number of uncondensed particles arising from thermal excitation\cite{Esko2011}. In order to quantify these properties, the depletion fraction of the particle number density at the core is defined as $\frac{\rho_{max}-\rho_{min}}{\rho_{max}}$. Here, $\rho_{min}$ is the density at the core, while $\rho_{max}$ is particle number density far away from core. The depletion fraction was investigated in Bogoliubov-de Gennes (BdG) theory in \cite{Pitaevskii2007}, where it was found that in the BEC regime the core contains nearly no particles, i.e. the depletion is almost  $100\%$, while in the BCS regime the soliton core still contains some particles at very low temperatures, with a depletion of about $30\%$. Furthermore, during the crossover from the BCS regime to the BEC regime, the depletion correspondingly undergoes a continuous variation from $30\%$ to  $100\%$.

Holographic methods can be used to describe unconventional superconductivity and superfluidity \cite{Hartnoll2008,Herzog2009}. Holographic dark solitons in the probe limit, i.e, neglecting the backreaction on the black-brane space-times forming the background, have been obtained in Refs.~\cite{Esko2009,Esko2010,Esko2011}. The depletion fraction for the particle number density was calculated in the two different quantizations possible in AdS/CFT, i.e, standard and alternative quantization. As explained in more detail below, both quantizations are related by a double-trace deformation. By comparing the holographic results with the depletion characteristics described above, a condensate with standard (alternative) quantization was found to show similar characteristics as in the BCS (BEC) regime in Ref.~\cite{Esko2011}. For ease of comparison, we will below refer to the condition of standard (alternative) quantization as BCS (BEC) case. Moreover, the authors of Refs.~\cite{Esko2010,Esko2011} conjectured that one may implement the BCS-BEC crossover in holographic superfluid systems by varying the scaling dimension of the condensing operator.

It was also argued in Ref.~\cite{DeWolfe2017} that by introducing double-trace deformations of the charged scalar, it is possible to model the interaction between fermionic constituents of the boundary system. By continuously tuning the coupling constant (denoted as $\kappa$ below) of the deformation, the salient features of the physics of the BCS-BEC crossover can be captured. In fact, the two independent arguments in Refs.~\cite{Esko2009,Esko2010,Esko2011} and Ref.~\cite{DeWolfe2017} are related: according to Refs.~\cite{Witten2001,Hartman2008, Vecchi2011,Johanna2013}, perturbing the large $\mathcal{N}$ boundary theory by a relevant double-trace deformation corresponds to imposing “mixed” boundary conditions $\psi_{+}=\kappa \psi_{-}$ for a scalar field in the bulk, where $\kappa$ is related to the coupling constant of the double-trace deformation.  $\kappa=0 $ ($\kappa=-\infty$) is the boundary conditions for alternative (standard) quantization. Adding the perturbation will trigger a renormalization group(RG) flow from the original conformal field theory(CFT) in the UV to another conformal fixed point in the IR. Under this flow, the conformal dimension of $\mathcal{O}$ varies between $\Delta_{-}$ in the UV and $\Delta_{+}$ in the IR, i.e,~the two inequivalent boundary CFTs $\Delta=\Delta_{\pm}$ can be recovered as two limits of the same deformed theory.

The argument of Refs.~\cite{Esko2009,Esko2010,Esko2011} leaves an important caveat: while the experimental results are obtained at nearly zero temperature, the holographic probe limit that ignores the backreaction of the matter fields onto the metric is known to break down in the low-temperature regime. In particular, the condensate in the alternative quantization diverges near zero temperature \cite{Hartnoll2008a}, which is a sign of the backreaction becoming important at low temperatures. In this work, we hence study the behavior of dark solitons in a holographic superfluid system including backreaction. The dark soliton configurations are constructed by numerically solving the Einstein equations coupled to the matter fields. In particular, we employ the DeTurck method for finding stationary solutions, first introduced in AdS/CFT in Refs.~\cite{Wiseman2010} and further developed in Ref.~\cite{Wiseman2011}, This method can reformulate the Einstein equations into a manifestly elliptic form by adding a covariant gauge-fixing term to the Einstein equations, which gives a well-posed boundary value problem.

Another drawback of the probe limit is that the boundary stress-energy tensor of the condensate cannot be investigated, which conceals information about important thermodynamic quantities such as the effective energy (mass) and entropy of the soliton. Taking into account the backreaction allows us to extract these quantities. Interestingly, it turns out that our result for the effective energy (mass) of the dark soliton, together with the surface tension of our holographic dark solitons, is consistent with previous expectations \cite{Fedichev1999,Cetoli2013} for the physical mechanism of a particular instability of dark solitons: the so-called \textit{snake instability}. The snake instability is an instability of solitons under transverse perturbations, leading to the spontaneous formation of a snake-like bending of the solitons. The snake instability was observed in different physical systems \cite{Anderson2001,Dutton2001,Yefsah2013}, and has attracted much theoretical attention \cite{Muryshev1999,Busch2000,Busch2001,Gangardt2010,Cetoli2013}. 
In holography, the authors of Refs.~\cite{Guo2020} identified the snake instability of holographic superfluids in the probe limit via the appearance of unstable quasinormal modes in the bulk, and observed the final decay of the ``snake'' into vortex-antivortex pairs. 
The investigation of Ref.~\cite{Guo2020} was systematic but not as intuitive as effective arguments from mechanics or hydrodynamics (see, e.g, Ref.~\cite{Fedichev1999,Cetoli2013}). In this work we holographically confirm the  explanation of the {\it snake instability} of dark solitons \cite{Fedichev1999,Cetoli2013} by calculating the negative effective mass responsible for the self-acceleration effect  \cite{Fedichev1999}, as well as the positive surface tension responsible for the spontaneous generation of ripples on the soliton \cite{Cetoli2013}.

This paper is organized as follows. In next section, we briefly introduce our  holographic superfluid model and analyze the \textit{Ans\"atze} and boundary conditions necessary for solving the equations of motion. In sec.~\ref{sec:numerics}, our numerical scheme and main numerical results are discussed. Sec.~\ref{sec:thermodynamics} is devoted to the thermodynamics of the holographic dark soliton and to holographically confirming the mechanism of the snake instability. A summary and outlook are included in Sec.~\ref{sec:conanddis}.

\section{Holographic Setup}

We work with the simplest holographic superfluid model which requires
gravity coupled to a Maxwell field $A_{\mu}$ and a massive charged scalar field
$\Psi$ with charge $e$. The bulk action reads
\begin{equation}
	S=\int d^{4}x\sqrt{-g}\left[\frac{1}{2\kappa_{4}^{2}}\left(R-2\Lambda\right)-\frac{1}{e^{2}}\left(\frac{1}{4}F_{\mu\nu}F^{\mu\nu}+\left|D\Psi\right|^{2}+m^{2}\left|\Psi\right|^{2}\right)\right],\label{eq:Action}
\end{equation}
where we have rescaled the gauge field $A_{\mu}$ and the scalar $\Psi$ to $\frac{A_{\mu}}{e}$ and $\frac{\Psi}{e}$
compared to the original form \cite{Hartnoll2008,Herzog2009}. $L$ is
the anti-de Sitter(AdS) radius related to the cosmological constant as $\Lambda=-\frac{3}{L^{2}}$, and $m$ is the mass of the charged scalar.
The covariant derivative is $D_\mu=\nabla_\mu-iA_\mu$, where $\nabla_\mu$ is the Christoffel covariant derivative with respect to the background metric $g_{\mu\nu}$. $F_{\mu\nu}=\partial_\mu A_\nu - \partial_\nu A_\mu$ is the field strength. In the rest of the paper we set $L=1$. The equations of motion derived from the action take the following form:

\begin{align}
	R_{\mu\nu}-\Lambda g_{\mu\nu}-\frac{2\kappa_{4}^{2}}{e^{2}}\left\{ \frac{1}{2}\left[D_{\mu}\Psi\left(D_{\nu}\Psi\right)^{\dagger}+D_{\nu}\Psi\left(D_{\mu}\Psi\right)^{\dagger}+g_{\mu\nu}m^{2}\left|\Psi\right|^{2}\right]\right.\nonumber \\
	\left.+\left(\frac{1}{2}F_{\mu\sigma}F_{\nu}^{\ \sigma}-\frac{g_{\mu\nu}}{8}F_{\rho\sigma}F^{\rho\sigma}\right)\right\} =0,\label{eq:EFT}
\end{align}

\begin{equation}
	D^{\mu}D_{\mu}\Psi-m^{2}\Psi=0,\label{eq:Scalar}
\end{equation}

\begin{equation}
	\nabla_{\mu}F^{\mu\nu}=ig^{\mu\nu}\left[\Psi^{\dagger}\left(D_{\mu}\Psi\right)-\Psi\left(D_{\mu}\Psi\right)^{\dagger}\right].\label{eq:Maxwell}
\end{equation}
In the probe limit $\frac{2\kappa_{4}^{2}}{e^{2}}\ll 1$, the backreaction of the terms involving the gauge field and the charged scalar on the background geometry in Eq.~\eqref{eq:EFT} can be neglected. In this limit, one can first solve the Einstein equations $R_{\mu\nu}-\Lambda g_{\mu\nu}=0$ for the fixed background metric $g_{\mu\nu}$, and then solve the matter equations \eqref{eq:Scalar} and \eqref{eq:Maxwell} on top of that fixed background. Once $\frac{2\kappa_{4}^{2}}{e^{2}}$ is not small this is no longer possible, and full coupled set of equations \eqref{eq:EFT}--\eqref{eq:Maxwell} have to be solved.  In this work we are interested in the effect of backreaction. In the following we will set the charge of the scalar $e=1$ 
and define the backreaction parameter $\epsilon\equiv2\kappa_{4}^{2}$ as a measure of the strength of backreaction.

In the absence of the charged scalar in Eq.~\eqref{eq:Action}, the
solution of the Einstein equations is the Reissner-Nordstr\"om-AdS
black brane
\begin{equation}\label{eq:AdSRN}
	ds^{2}=\text{\ensuremath{\frac{1}{z^{2}}\left[-f\left(z\right)dt^{2}+\frac{dz^{2}}{f\left(z\right)}+dz^{2}+dy^{2}\right]}},
\end{equation}
\begin{align}
	f\left(z\right) & =1-\left(1+\frac{\epsilon\mu^{2}z_{+}^{2}}{4}\right)\left(\frac{z}{z_{+}}\right)^{3}+\frac{\epsilon\mu^{2}z_{+}^{2}}{4}\left(\frac{z}{z_{+}}\right)^{4},
\end{align}
\begin{equation}
	A=\mu\left[1-\left(\frac{z}{z_{+}}\right)\right]dt,
\end{equation}
where $\mu$ is the chemical potential and $z_{+}$ parametrizes
the black-brane temperature via 
\begin{equation}
	T=\frac{1}{4\pi z_{+}}\left(3-\frac{\epsilon\mu^{2}z_{+}^{2}}{4}\right)\,.\label{eq:Hawking temperature}
\end{equation}
For numerical convenience \cite{Ren2017}, we make the following radical
coordinate transformation with $z_{h}\equiv\frac{1}{z_{+}}$:
\begin{equation}\label{rcoord}
	z=\frac{1-r^{2}}{z_{h}}.
\end{equation}

In order to construct the backreacted geometries, we use the following \textit{Ansatz} compatible with staticity and translation invariance in the second boundary direction $y$:
\begin{equation}
	ds^{2}=\text{\ensuremath{\frac{z_{h}^{2}}{\left(1-r^2\right)^{2}}\left[-Q_{1}f\left(r\right)dt^{2}+\frac{4r^{2}Q_{2}dr^{2}}{z_{h}^{2}f\left(r\right)}+Q_{4}\left(dx-\frac{2r}{z_{h}}Q_{3}dr\right)^{2}+Q_{5}dy^{2}\right]}},\label{eq:Ansatz}
\end{equation}
\begin{equation}
	\Psi=\left(\frac{1-r^{2}}{z_{h}}\right)Q_{6},
\end{equation}
\begin{equation}
	A=\mu r^{2}Q_{7}dt\,.
\end{equation}
Here ${\left\{  Q_{i}|i=1,2,\cdots,7\right\} }$
are functions of $r$ and $x$ to be determined by solving Eqs.~\eqref{eq:EFT}--\eqref{eq:Maxwell}. 
In the coordinate \eqref{rcoord}, the conformal boundary is located at $r=1$, while the horizon is at $r=0$ where the regularity of $Q_{i}$ must be imposed. Expanding the equations of motion near the horizon as a power series in $r$ and requiring the leading order to vanish yields the boundary conditions
\begin{equation}
	Q_{1}|_{r=0}=Q_{2}|_{r=0};\left(\partial_{r}Q_{j}\right)|_{r=0}=0,j=2,3,\cdots7.\label{eq:bcsr0}
\end{equation}
In particular, the Dirichlet condition for $Q_1$  in Eq.~\eqref{eq:bcsr0}  ensures that the temperature of the black brane is still given by Eq.~\eqref{eq:Hawking temperature}. At the conformal boundary, we demand that the metric approaches AdS$_{4}$, i.e. 
\begin{equation}
	Q_{1}|_{r=1}=Q_{2}|_{r=1}=Q_{4}|_{r=1}=Q_{5}|_{r=1}=1;Q_{3}|_{r=1}=0.\label{eq:bcsr1}
\end{equation}
In the asymptotically AdS regime, the scalar field $\Psi$ behaves in the $z$ coordinate as
\begin{equation}
	\Psi=\psi_{-}z^{\Delta_{-}}+\psi_{+}z^{\Delta_{+}}+\ldots
\end{equation}
Here $\Delta_{\pm}=3/2\pm\sqrt{9/4+m^{2}L^{2}}$ is the conformal dimension of the operator $\mathcal{O}$ dual to $\psi$. We  choose $m^2L^2=-2$.  According to the AdS/CFT correspondence, in the standard quantization, one  identifies $\psi_{-}$ with the source of the operator $\mathcal{O}_{\text{\ensuremath{2}}}$ having the conformal dimension $\Delta_{+}=2$, while $\psi_{+}$ is regarded as the response (expectation value). Alternatively, one can exchange the roles of $\psi_{-}$ and $\psi_{+}$ in the dual field theory, i.e,  $\psi$ is dual to  operator $\mathcal{O}_{1}$ having the conformal dimension $\Delta_{-}=1$ with source $\psi_{+}$ and expectation value $\psi_{-}$. Since we want to investigate a spontaneously generated condensate in the absence of sources, we will set $\psi_{-}=0$ (standard case) or $\psi_{+}=0$ (alternative case) on the conformal boundary. Generally, the mixed boundary condition $\psi_{+}=\kappa\psi_{-}$ can also be imposed, which is known as double-trace boundary condition since it corresponds to a deformation of the dual theory by adding a term $\sim\int d^{3}x\mathcal{O}^{\dagger}\mathcal{O}$ to its boundary action S$_{bdry}$ \cite{Witten2001,Hartman2008, Vecchi2011}. Finally, the gauge field admits the usual UV expansion
\begin{equation}
	A_{t}=\mu-\rho z+\ldots
\end{equation}
$\mu$ is interpreted as  the chemical potential, and $\rho$ is the charge (or particle number) density. Since the ground state of the system is conformal, we can scale out one dimensionful quantity. Throughout this paper, we do so by normalizing all dimensionful quantities with the chemical potential, which we set to the fixed value $\mu=5.6$.\footnote{This value turns out to be numerically convenient in terms of the convergence speed of our code.}

\section{Numerical Scheme and Results\label{sec:numerics}}

We employ the DeTurck method to numerically solve the Einstein equations, for a recent review, see Ref.~\cite{Santos2016}. This method consists of adding the gauge-fixing term $\frac{1}{2}\mathcal{L_{\xi}}g_{\mu\nu}=\nabla_{(\mu}\xi_{\nu)}$ to the Einstein equations \eqref{eq:EFT}, which breaks all diffeomorphisms and yields elliptic  Einstein-DeTurck equations,
\begin{align}
	R_{\mu\nu}-\Lambda g_{\mu\nu}-\epsilon\left\{ \frac{1}{2}\left[D_{\mu}\Psi\left(D_{\nu}\Psi\right)^{\dagger}+D_{\nu}\Psi\left(D_{\mu}\Psi\right)^{\dagger}+g_{\mu\nu}m^{2}\left|\Psi\right|^{2}\right]\right.\nonumber \\
	\left.+\left(\frac{1}{2}F_{\mu\sigma}F_{\nu}^{\ \sigma}-\frac{g_{\mu\nu}}{8}F_{\rho\sigma}F^{\rho\sigma}\right)\right\} -\nabla_{(\mu}\xi_{\nu)}=0.\label{eq:Deturck}
\end{align}
Here the DeTurck vector $\xi^{\nu}\equiv g^{\rho\sigma}\left[\text{\ensuremath{\Gamma_{\rho\sigma}^{\nu}}\ensuremath{\left(g\right)}}-\text{\ensuremath{\overline{\Gamma}_{\rho\sigma}^{\nu}}\ensuremath{\left(\bar{g}\right)}}\right]$
is constructed from the difference of the Christoffel symbols of the metric $g_{\mu\nu}$ (which we aim to solve for) and a  reference metric $\bar{g}_{\mu\nu}$. The reference metric has to have the same
asymptotics and symmetries as the metric $g_{\mu\nu}$ we are trying to solve for. In our scheme, we take the standard Reissner-Nordstr\"om-AdS metric \eqref{eq:AdSRN}, corresponding to
$Q_{1}=Q_{2}=Q_{4}=Q_{5}=1$ and $Q_{3}=0$ in Eq.~\eqref{eq:Ansatz}. We then find solutions to the Einstein-DeTurck equation with the constraint condition $\xi^\mu=0$, which ensures that our solution coincides with a solution to Einstein equations \eqref{eq:EFT}.\footnote{We have checked  that $|\xi^2| < 10^{-10}$ when the size of the grids reaches $30\times150$.}

The nonlinear partial differential equations(PDEs) \eqref{eq:Scalar}, \eqref{eq:Maxwell}, and \eqref{eq:Deturck}, together with the boundary conditions \eqref{eq:bcsr0}--\eqref{eq:bcsr1} are then solved via the Newton-Kantorovich method. To be specific, we first  linearize the PDEs and then discretize the linear partial differential equations into algebraic equations via the standard pseudospectral procedure, where we represent unknown functions as linear combinations of Chebyshev polynomials in the $z$ coordinate and a Fourier series in the $x$ coordinate.\footnote{Since a single soliton has no periodicity in the \emph{x} direction, for the simplicity of the spatial boundary conditions and efficiency of the numerics, we follow Ref.~\cite{Lan2017} and instead construct the double soliton (kink-antikink) configuration, and then analyze a single soliton.}  Our integration domains lives on a rectangular grid, $\left(r,x\right)\in\left(0,1\right)\times\left(-\frac{L_{x}}{2},\frac{L_{x}}{2}\right)$.
The resulting linear system is solved by lower-upper decomposition or other iterative techniques.
\begin{figure}
	\centering
	\hfill{}\subfloat[Critical temperature]{\includegraphics[width=0.43\columnwidth]{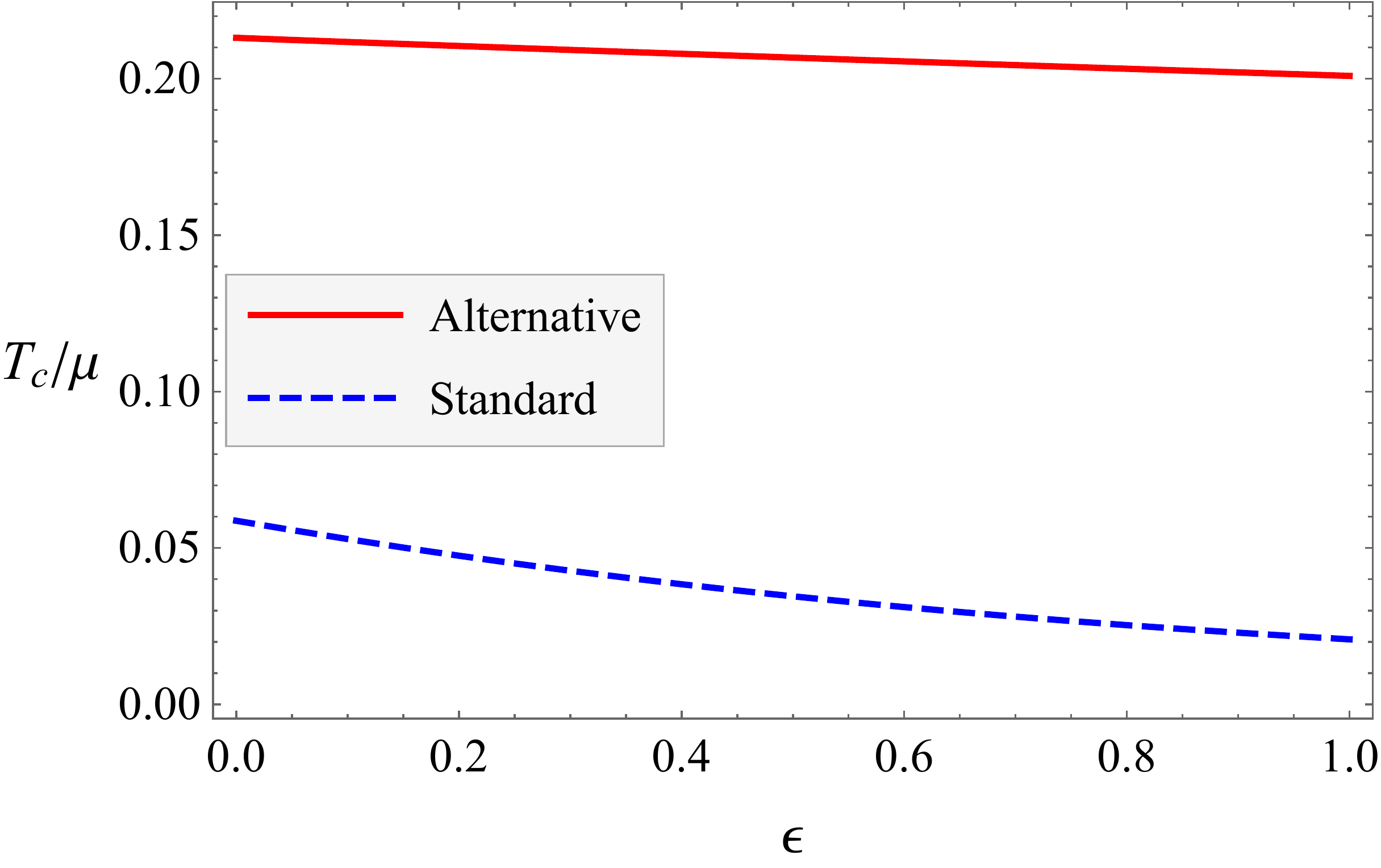}
		
	}\hfill{}\subfloat[Normalized critical temperature]{\includegraphics[width=0.43\columnwidth]{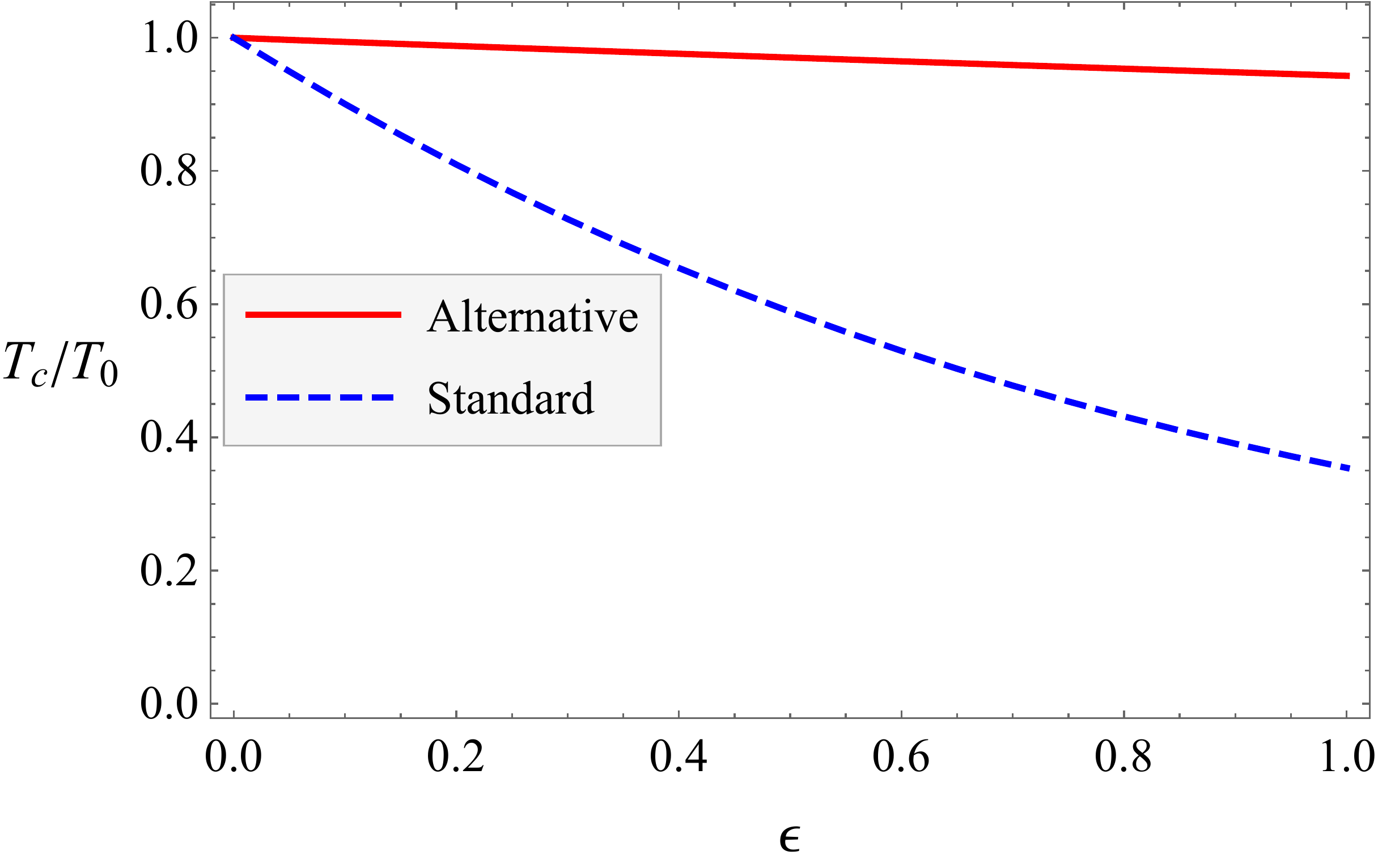}}\hfill{}
	
	\caption{\label{fig:critical temperature}Absolute (a) and normalized (b) critical temperature for varying backreaction parameter $\epsilon$. $T_{0}$ is the critical temperature in the probe limit $\epsilon=0$. The critical temperature drops with increasing backreaction for both the standard and alternative quantization, indicating a suppression of the condensate mechanism.}
\end{figure}

At a critical temperature $T=T_{c}$, an instability of the black brane to form charged scalar hair occurs. This instability corresponds to a phase transition in the boundary theory. We begin by a discussion of the critical temperature $T_c$ and its dependence on the backreaction as given by the backreaction parameter $\epsilon  = 2\kappa_{4}^{2}$. To determine where our solutions become unstable, we have to determine the critical temperature for a given $\epsilon$. We do this by perturbing the Reissner-Nordstr\"om-AdS \eqref{eq:AdSRN} with the scalar field $\psi=\phi\left(r\right)e^{-i\omega_{p} t}$. At the onset of the instability, the frequency of the unstable mode crosses zero,  $\omega_{p} = 0$. The critical temperature $T_{c}$ itself is therefore found by looking for a static normalizable solution to the scalar equation of motion with $\omega_{p}=0$. The results are shown in Fig.~\ref{fig:critical temperature}. We find that in alternative quantization,  $T_c/\mu$ is always higher than in the standard case, since the alternative boundary conditions allow for more modes to become unstable in the IR, i.e.~in the deep interior of the space,  in the sense that they violate the Breitenlohner-Freedman bound there \cite{Horowitz2011}. Furthermore, in AdS/CFT the mass of the scalar field in the bulk determines the dimension of the the operator in standard ($\Delta_+$)/alternative ($\Delta_-$) quantization via $\Delta_\pm = \frac{d}{2} \pm \sqrt{\frac{d^2}{4} + m^2 L^2}$. From the field theoretic point of view, the dimension of the operator in alternative quantization is always smaller than in the standard case, $\Delta_- < \Delta_+$.\footnote{Equality is reached at the Breitenlohner-Freedman bound $m^2L^2 = - d^2/4$.} The operator with smaller scaling dimension is more relevant in the RG sense, and its effects will be more pronounced already at larger values of the renormalization scale. This operator hence will condense at higher temperatures, explaining the larger $T_c$ in the alternative case.  

In relation to this, we find that for for a value of the backreaction parameter in the range $\epsilon\in[0,1]$, the critical temperature for the alternative (BEC) case has only a milder dependence on $\epsilon$ compared to the standard (BCS) case. Within this range, the backreaction is relatively weak, and we can expand the blackening factor  $f(z)$ in a power series in $\epsilon$ near the probe limit $\epsilon=0$ \cite{Hartnoll2008,Hartnoll2008a},\footnote{A possible hyperscaling violating factor has been ruled out by the construction of the IR fixed point in  \cite{Horowitz2011}.} with the first correction being 
\begin{align}
f\left(z\right)  =1-\left(\frac{z}{z_{+}}\right)^{3}+\epsilon\left(\frac{3\mu}{8\pi T}\right)^2\left(\frac{z}{z_{+}}\right)^{3}\left(\frac{z}{z_{+}}-1\right)+\dots.
\end{align}
Since $\epsilon$ is multiplied by $(\mu/T)^2$, just below the critical temperature the effect of backreaction is suppressed for higher $T_c/\mu$. Since $T_c/\mu$  is at least a factor 4 larger in the alternative case, the backreaction-induced change on $f(z)$ in  alternative quantization is reduced by a factor of 16 as compared to the standard case. This implies that in the alternative case, the equations of motions are far less sensitive to changes in the backreaction parameter. In particular the  change in $T_c/\mu$ to order $\epsilon$ is also going to be suppressed compared to the standard case. In summary, the insensitivity of $T_c/\mu$ on the backreaction parameter in the alternative case is due to the enhancement of $T_c/\mu$ in the probe limit as compared to the standard case.

\begin{figure}
	\includegraphics[width=0.6\linewidth]{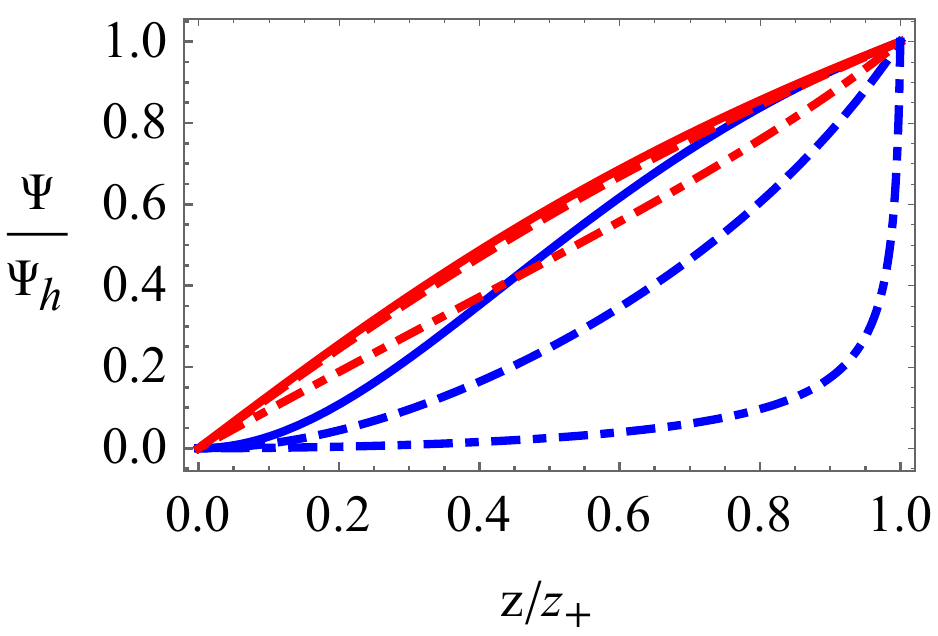}
	\caption{Profiles of $\Psi/\Psi_h$ just below the critical temperature, where $\Psi_h$ is the value of $\Psi$ on the horizon. We show the cases of $\epsilon=0$ (solid line), $\epsilon=1$ (dashed line) and $\epsilon=10$ (dot-dashed line). The blue (red) line refers to the profile in the standard (alternative) quantization.}
	\label{Scalar}	
\end{figure}
\begin{figure}
	\includegraphics[width=0.6\linewidth]{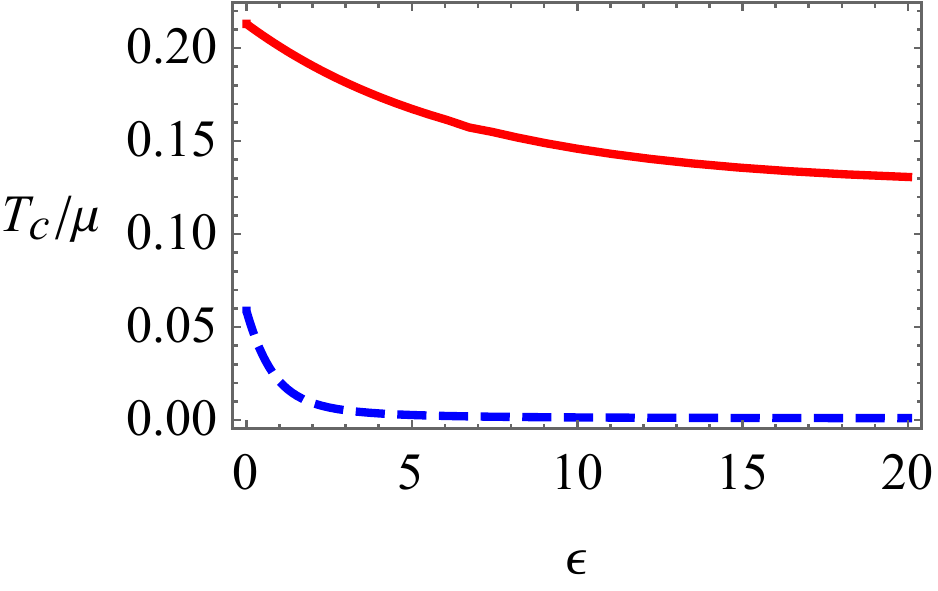}
	\caption{$T_c/\mu$ in the standard quantization (blue) and alternative quantization (red).}
	\label{FullTc}	
\end{figure}
The previous argument is corroborated by the change in the scalar field profile $\Psi$  as shown for the scalar in Fig.~\ref{Scalar}, which is also found to be much weaker in alternative quantization compared to the standard case. Also the boundary condition of $\Psi$ on the horizon, 
\begin{align}
\left.\frac{\partial_\xi\Psi}{\Psi}\right|_{\xi=1}=-\frac{8}{\epsilon(3\mu/4\pi T)^2 -12}+\dots,\quad \xi=\frac{z}{z_+},
\end{align}
depends more mildly on $\epsilon$ in alternative quantization. Thus we see that in all cases $T_c/\mu$ in the probe limit controls the first correction in the backreaction parameter $\epsilon$, and hence in alternative quantization depends more weakly on the backreaction. However, in a wider range of $\epsilon$, the dependence on $\epsilon$ can only be found numerically, as shown in Fig.~\ref{FullTc}.

\begin{figure}
	\centering
	\hfill{}\subfloat[$Q_1$]{\includegraphics[scale=0.25]{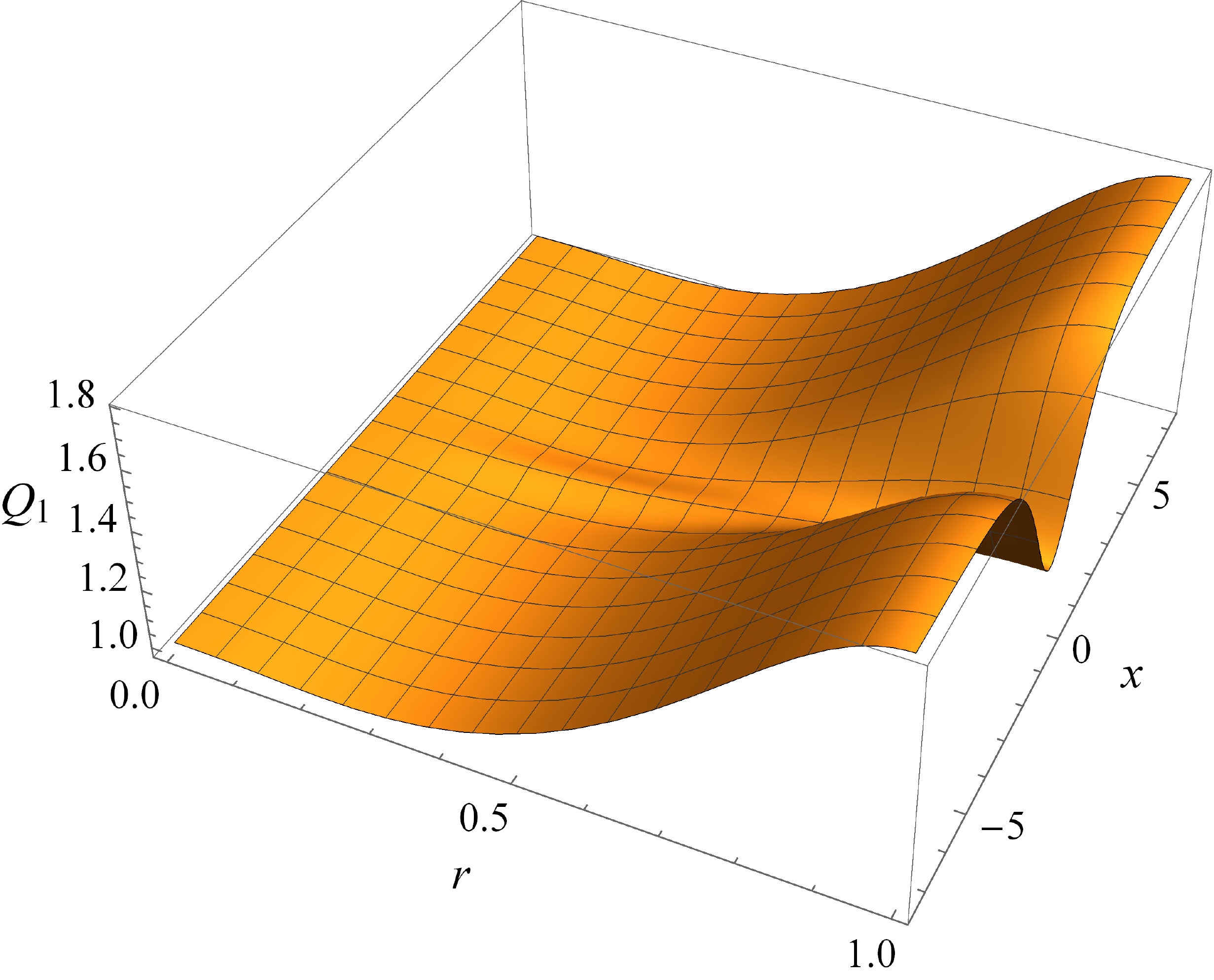}}\hfill{}\subfloat[$Q_3$]{\includegraphics[scale=0.25]{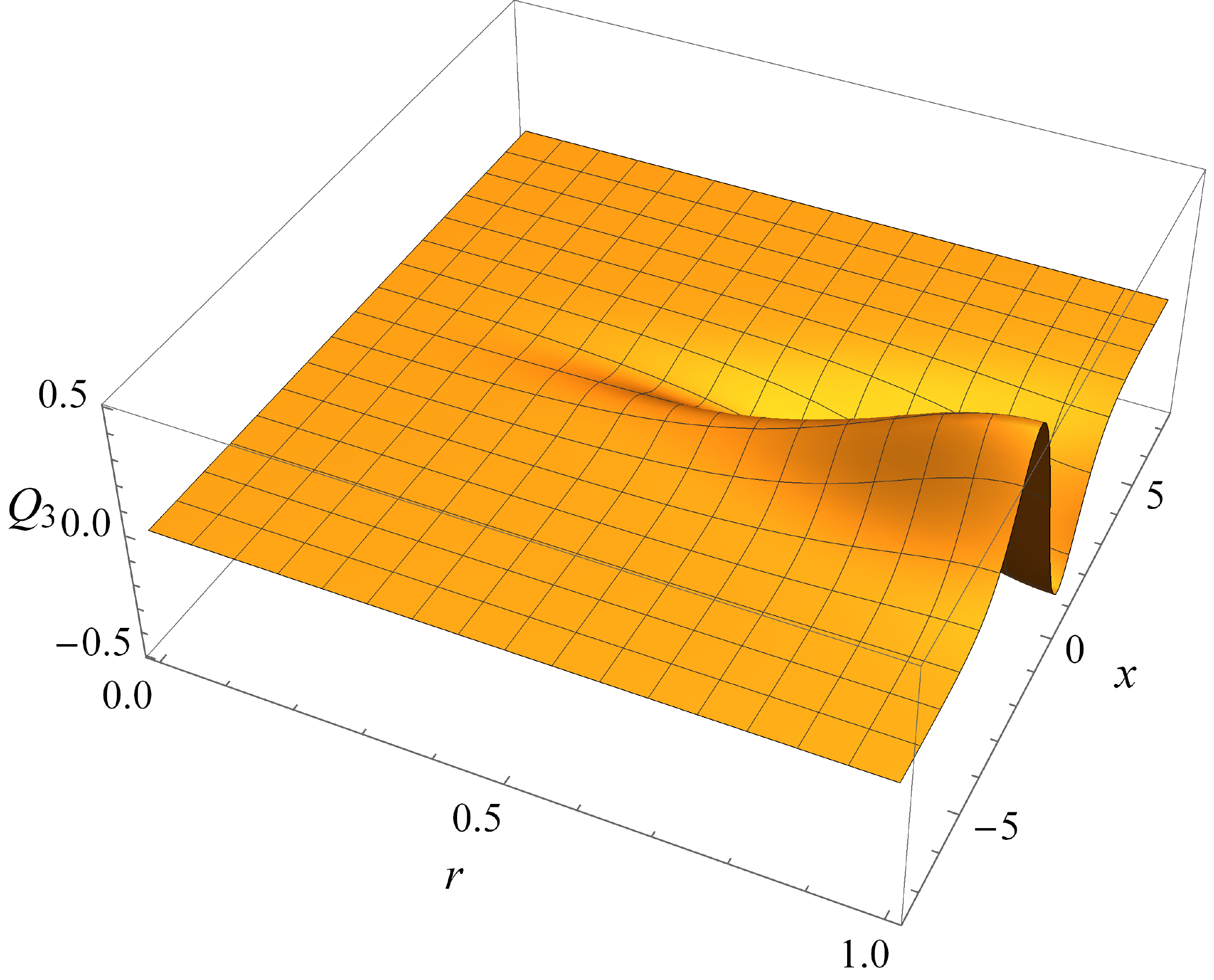}}\hfill{}
	
	\hfill{}\subfloat[$Q_6$]{\includegraphics[scale=0.25]{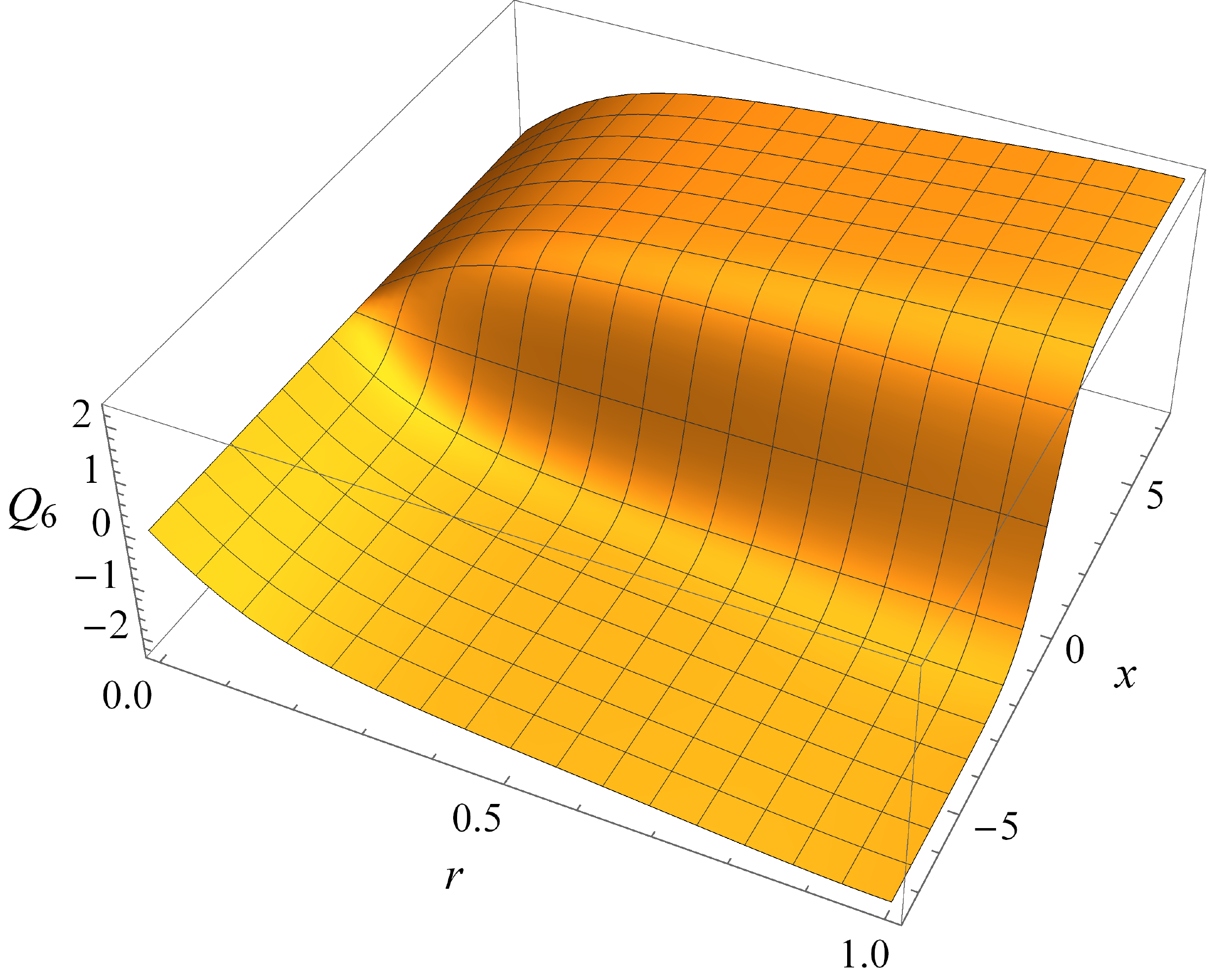}}\hfill{}\subfloat[$Q_7$]{\includegraphics[scale=0.25]{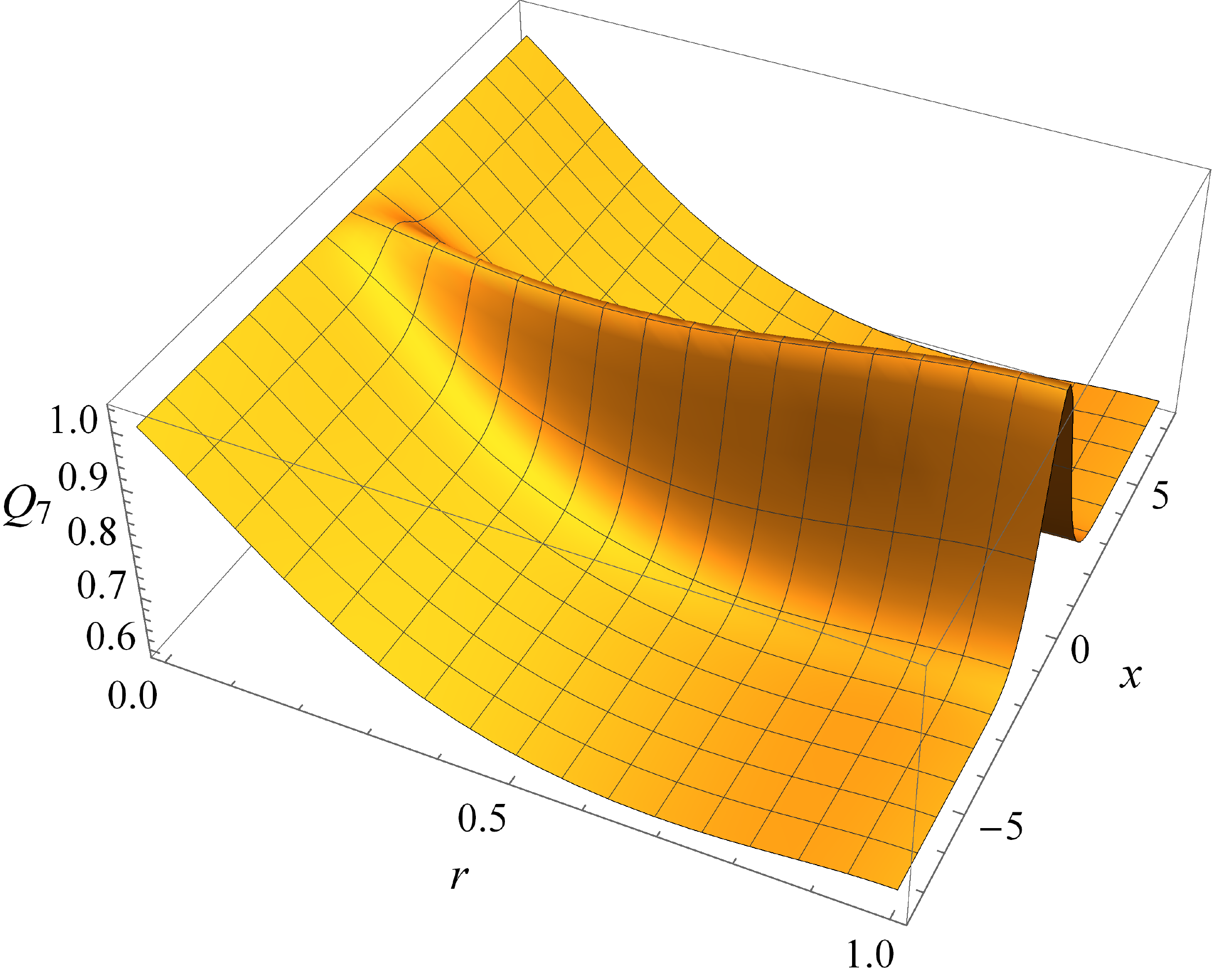}}\hfill{}
	\caption{\label{fig:profileO2}The profile of (a) the function $Q_1$ setting the $tt$ metric component, (b) the function $Q_3$ setting, together with $Q_4$, the $rr$ metric component, (c) the function $Q_6$ setting the charged scalar profile, and (d) the function $Q_7$ setting the $t$ component of the gauge field, at $\epsilon=0.25,T/T_{c}=0.5$ in the standard quantization.}
\end{figure}

\begin{figure}
\centering
\hfill{}\subfloat[$Q_1$]{\includegraphics[scale=0.25]{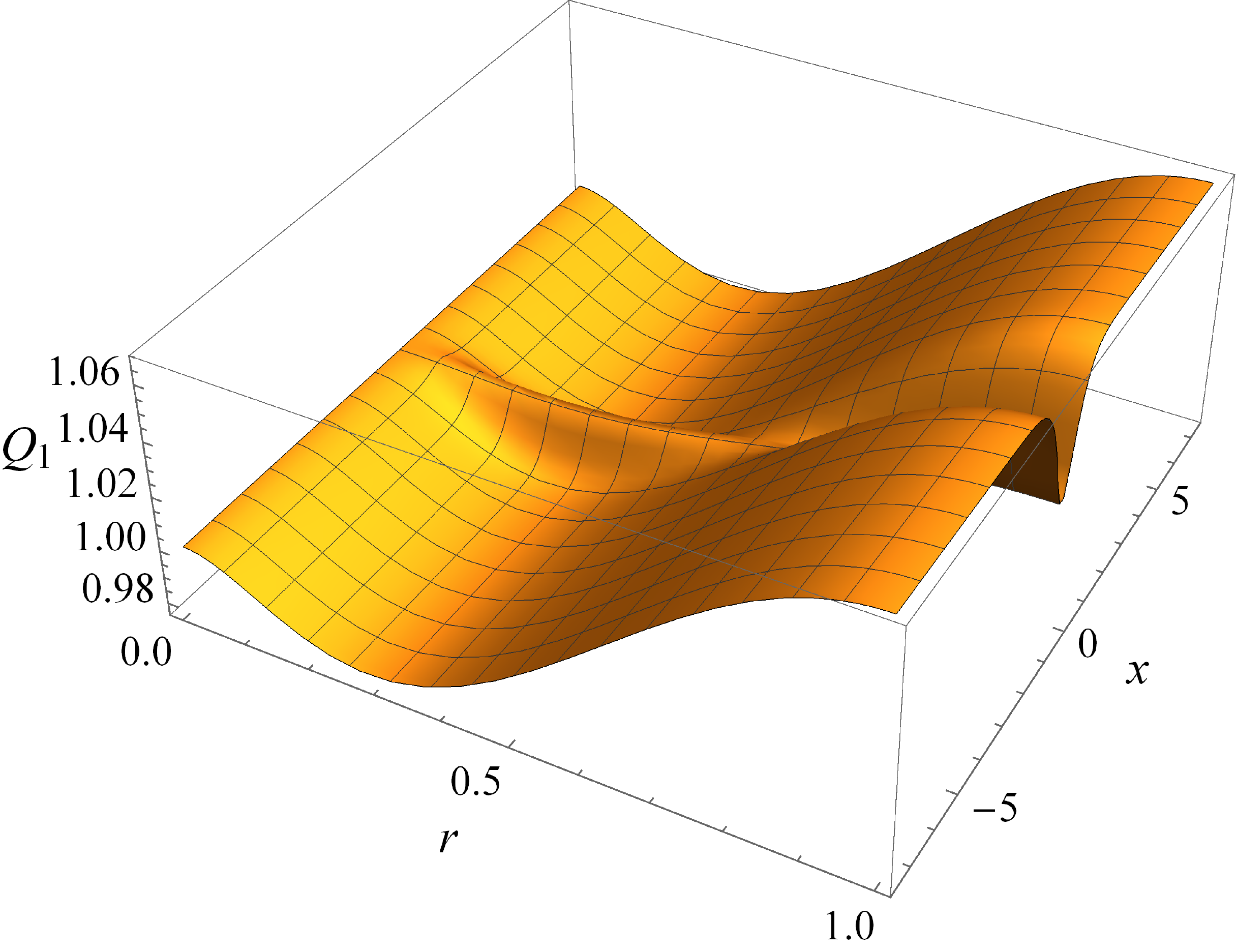}}\hfill{} \subfloat[$Q_3$]{\includegraphics[scale=0.25]{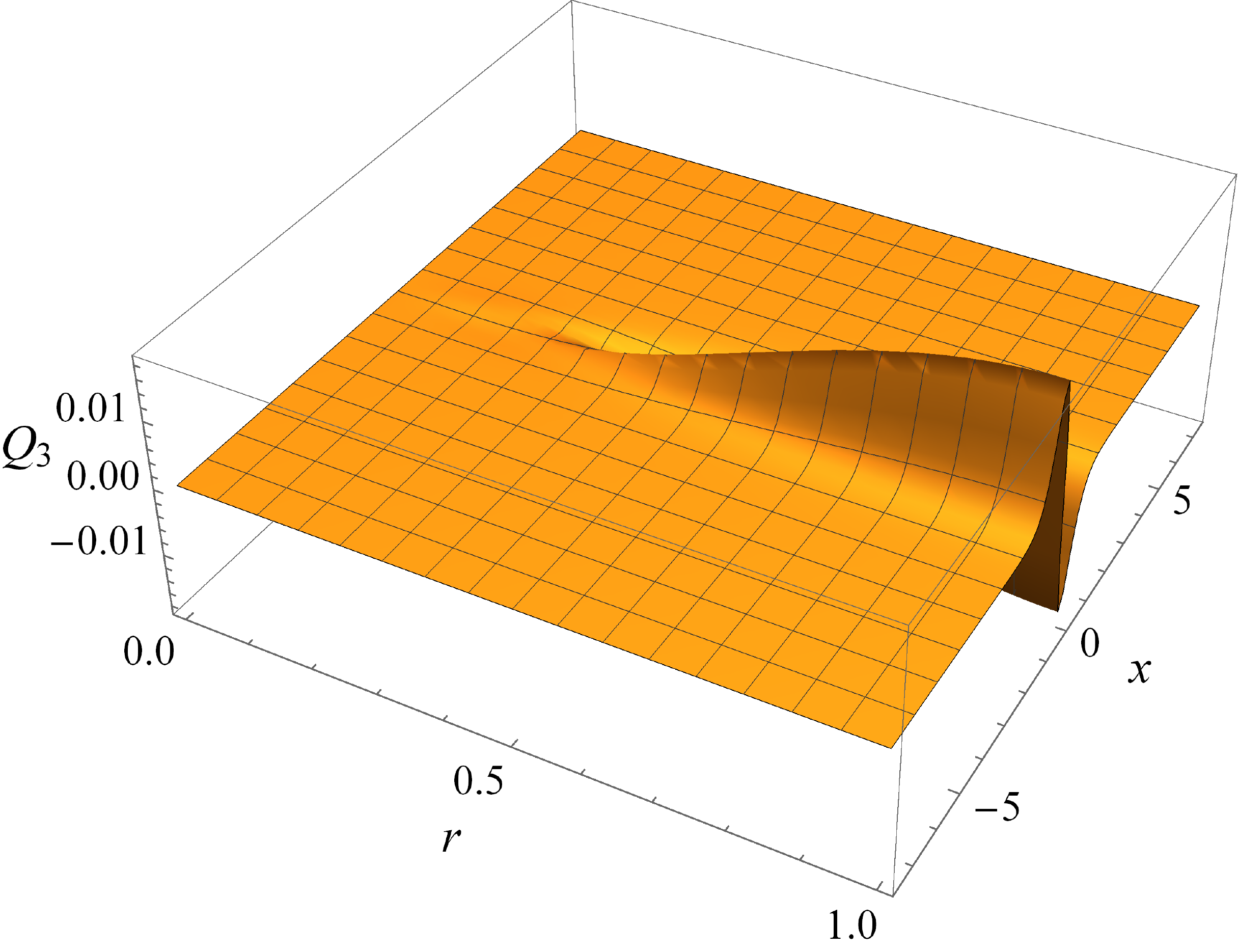}}\hfill{}

\hfill{}\subfloat[$Q_6$]{\includegraphics[scale=0.25]{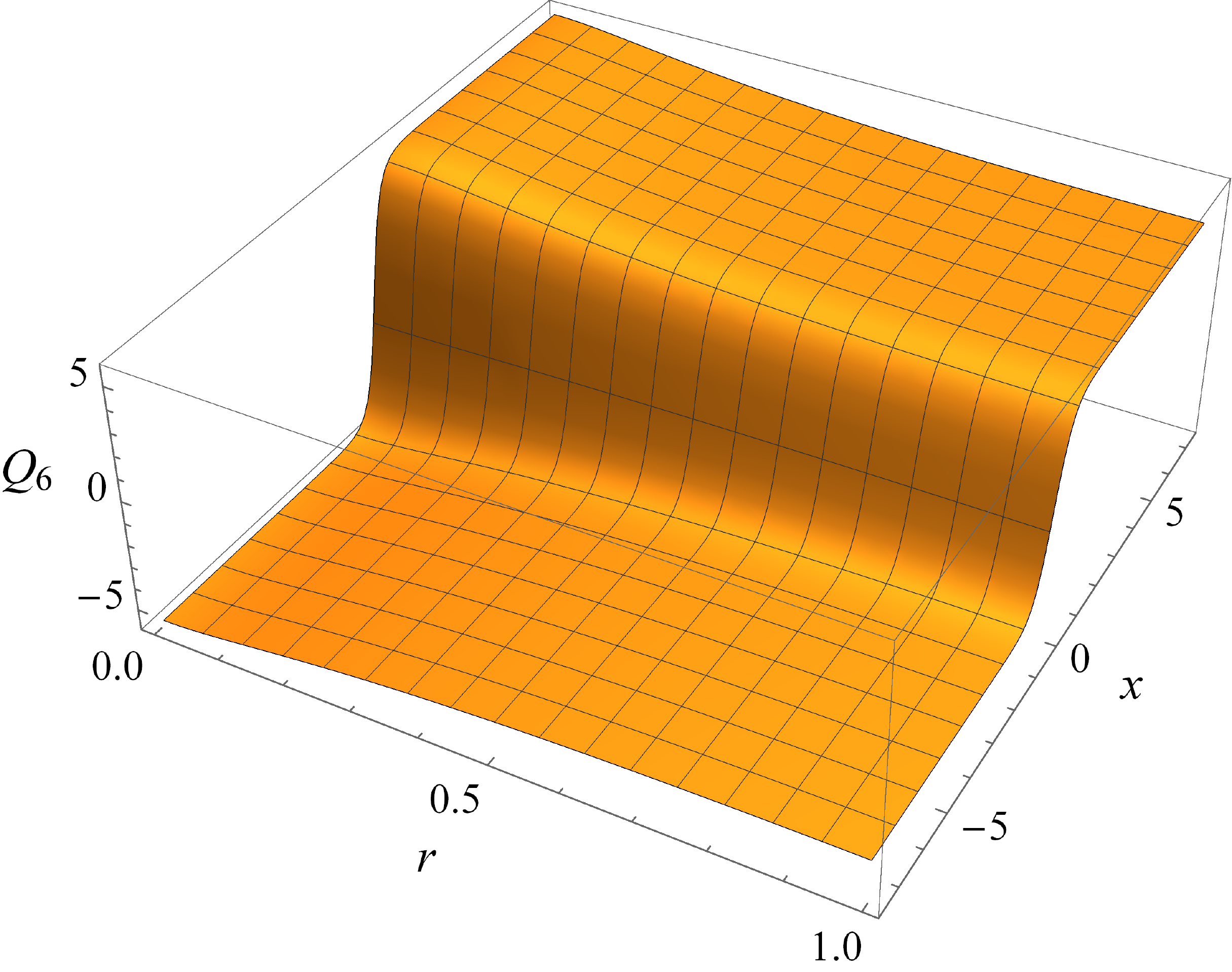}}\hfill{} \subfloat[$Q_7$]{\includegraphics[scale=0.25]{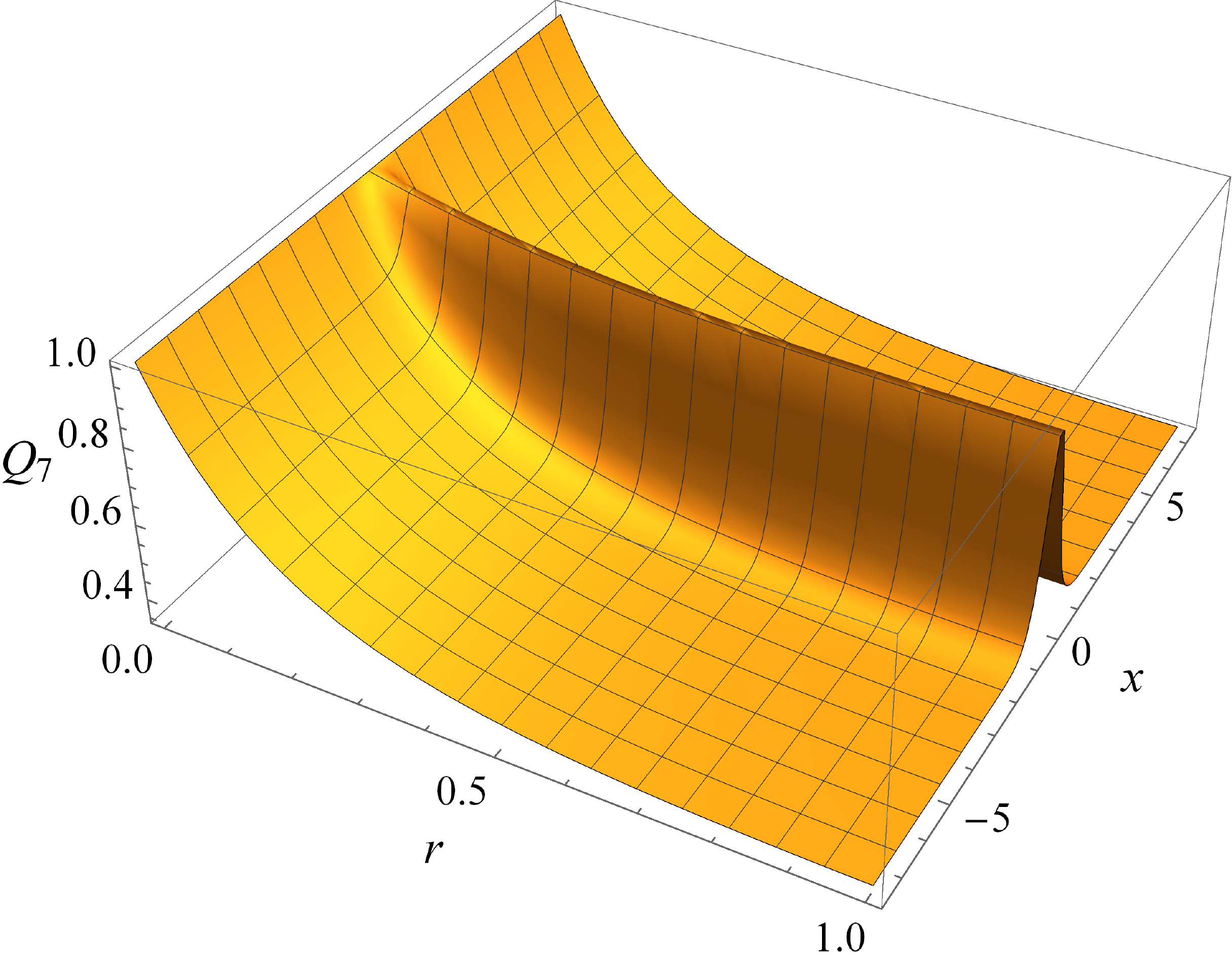}}\hfill{}
\caption{\label{fig:profileO1}The profile of (a) the function $Q_1$ setting the $tt$ metric component, (b) the function $Q_3$ setting, together with $Q_4$, the $rr$ metric component, (c) the function $Q_6$ setting the charged scalar profile, and (d) the function $Q_7$ setting the $t$ component of the gauge field, at $\epsilon=0.25,T/T_{c}=0.5$ in the alternative quantization.}
\end{figure}

Based on the above results for the critical temperature, we now construct the solution with backreaction for $T<T_{c}$. These are hairy charged black holes dual to the superfluid phase. The seed configurations of the matter field can be chosen as $\psi_{\pm} \sim \tanh(x-L_{x}/4)\tanh(-x-L_{x}/4)$. As a result, the components of metric and the configuration of  the matter fields are shown in Figs.~\ref{fig:profileO2} and~\ref{fig:profileO1}. We can see that larger fluctuations of the spacetime metric appear only near the core of the soliton.

\begin{figure}
\centering
\subfloat[Condensate (order parameter)]{\includegraphics[width=0.43\columnwidth]{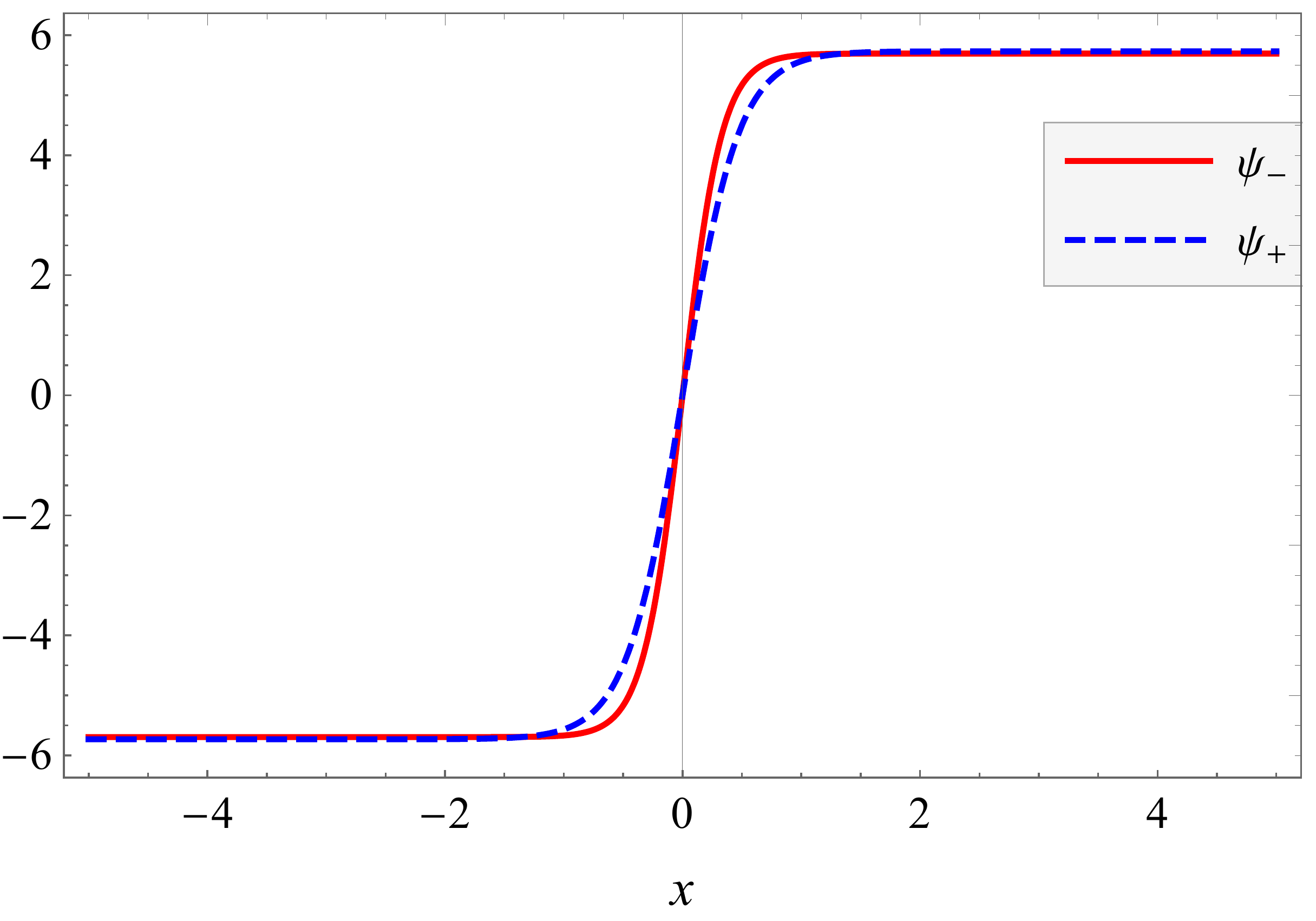}}\quad \subfloat[Particle number density]{\includegraphics[width=0.43\columnwidth]{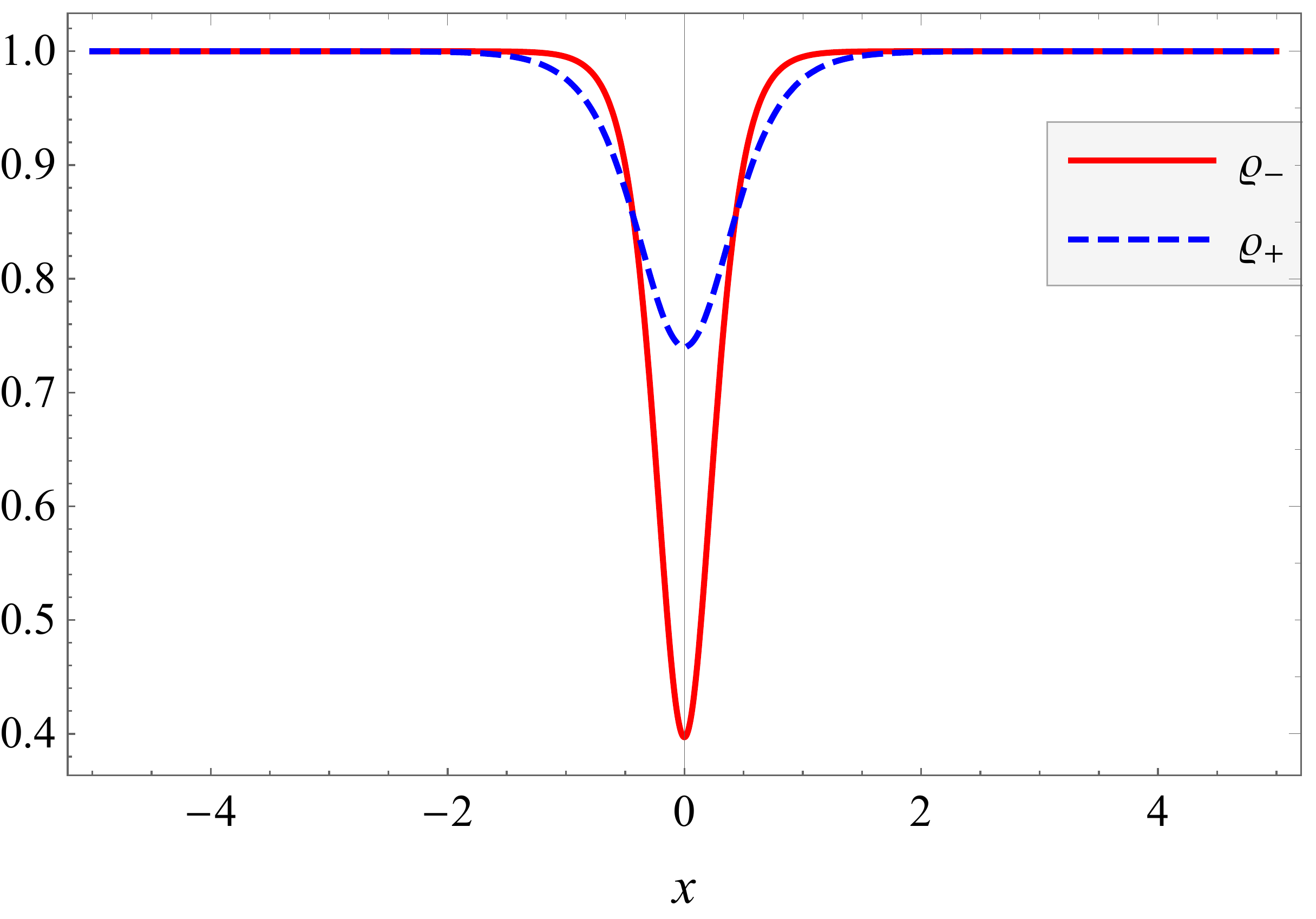}}\caption{\label{fig:condensate and density}  The condensate (a) and  the particle
number density (b) as a function of $x$ at $\epsilon=0.25,T/T_{c}=0.5$.
$\varrho_{+}$ and $\varrho_{-}$ are particle number
densities normalized with their equilibrium values at $x\rightarrow \pm \infty$. Red solid lines correspond to the alternative (BEC) case, while blue dashed lines correspond to the standard (BCS) case. The charge depletion is considerably larger in the BEC case compared to the BCS case.}
\end{figure}

\begin{figure}
\centering
\hfill{}\subfloat[Standard (BCS) case]{\includegraphics[width=0.43\columnwidth]{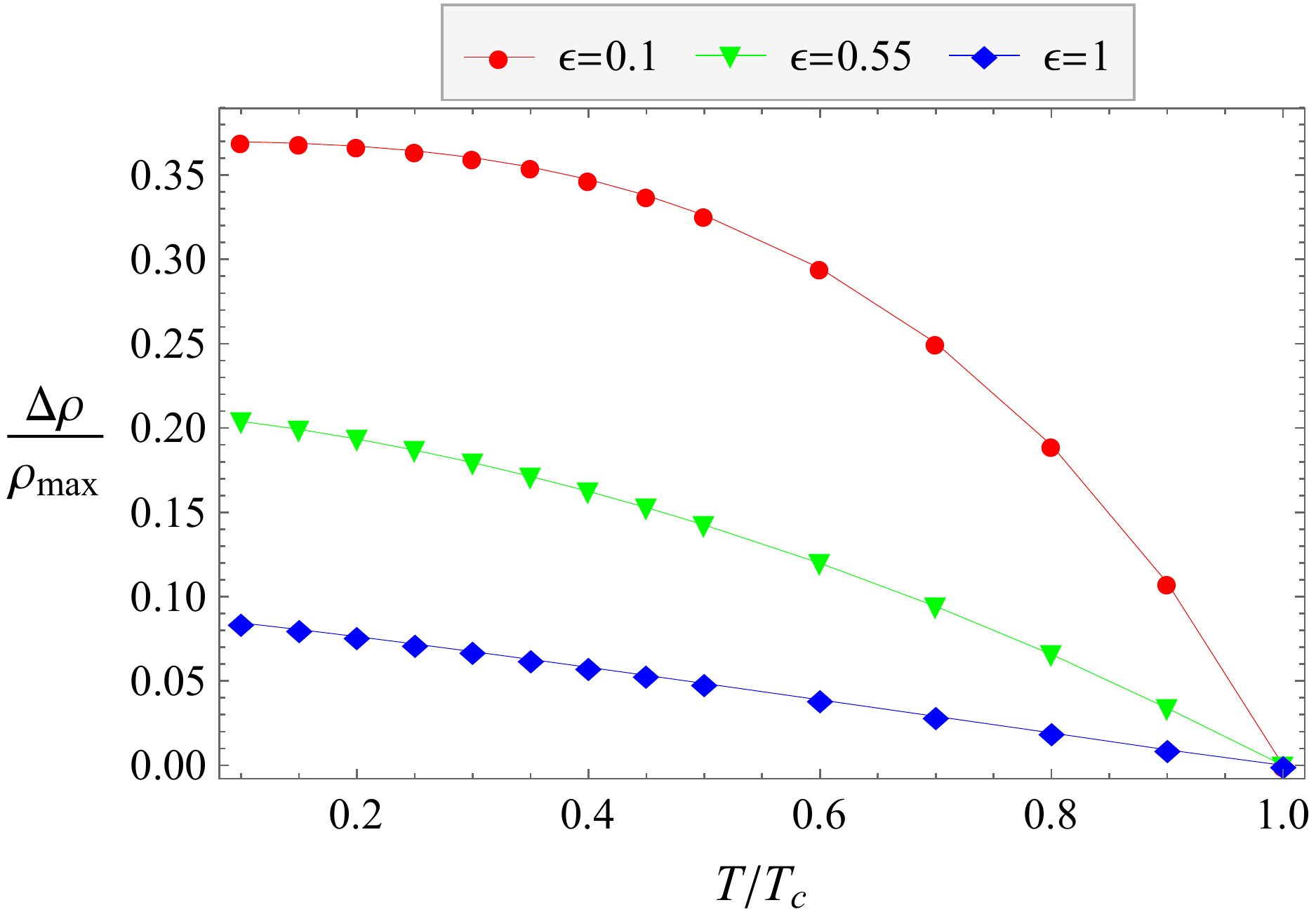}}
\hfill{}\subfloat[Alternative (BEC) case]{\includegraphics[width=0.43\columnwidth]{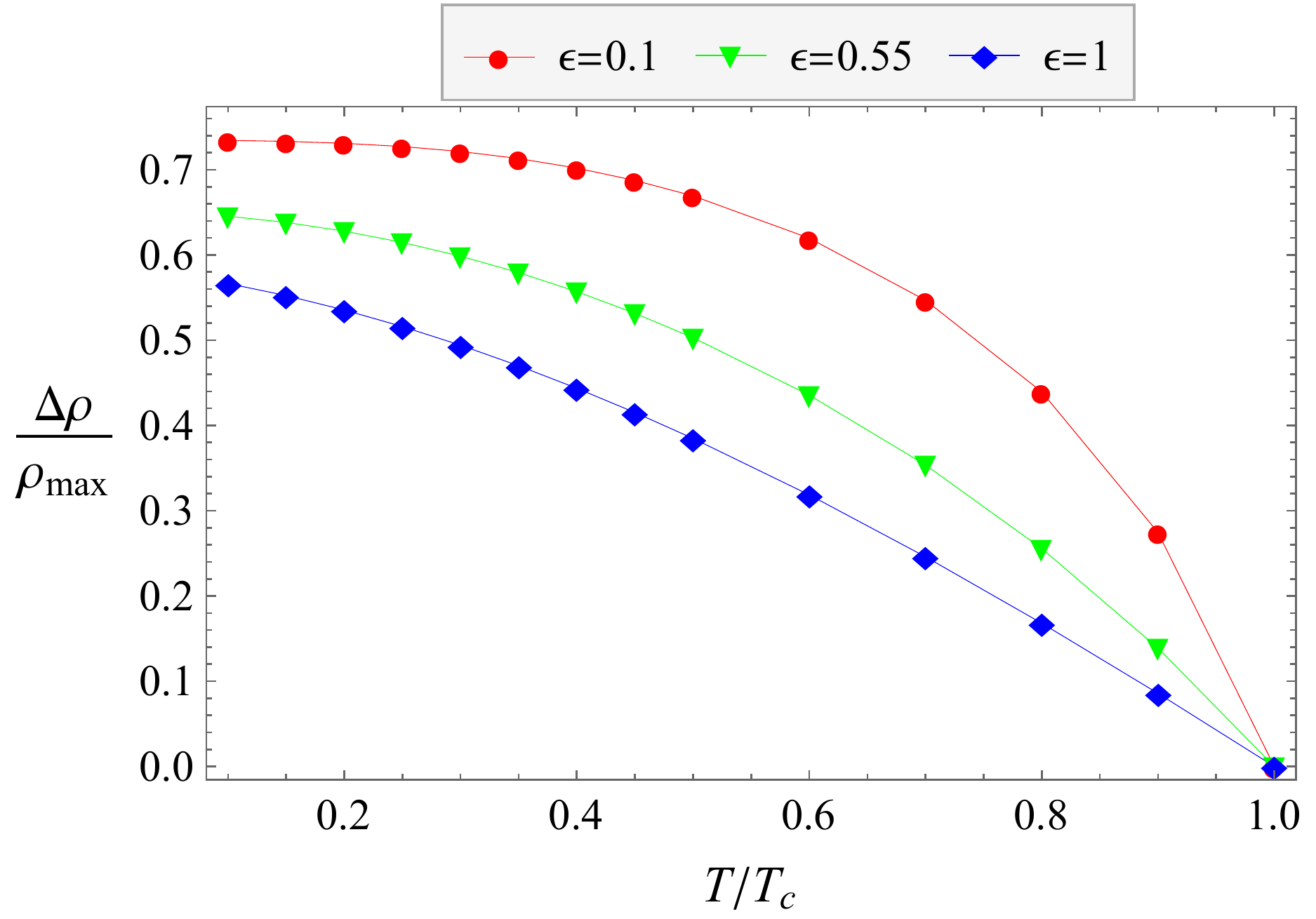}}\hfill{}
\caption{\label{fig:depletionvsT}The depletion of particle number density
as a function of $T/T_{c}$ for different $\epsilon$ in the (a) standard (BCS) and (b)  alternative (BEC) case. In both cases, the depletion factor decreases with increasing backreaction, i.e, more charge is present at the soliton core.}
\end{figure}

\begin{figure}
\centering
\hfill{}\subfloat[Standard (BCS) case]{\includegraphics[width=0.43\columnwidth]{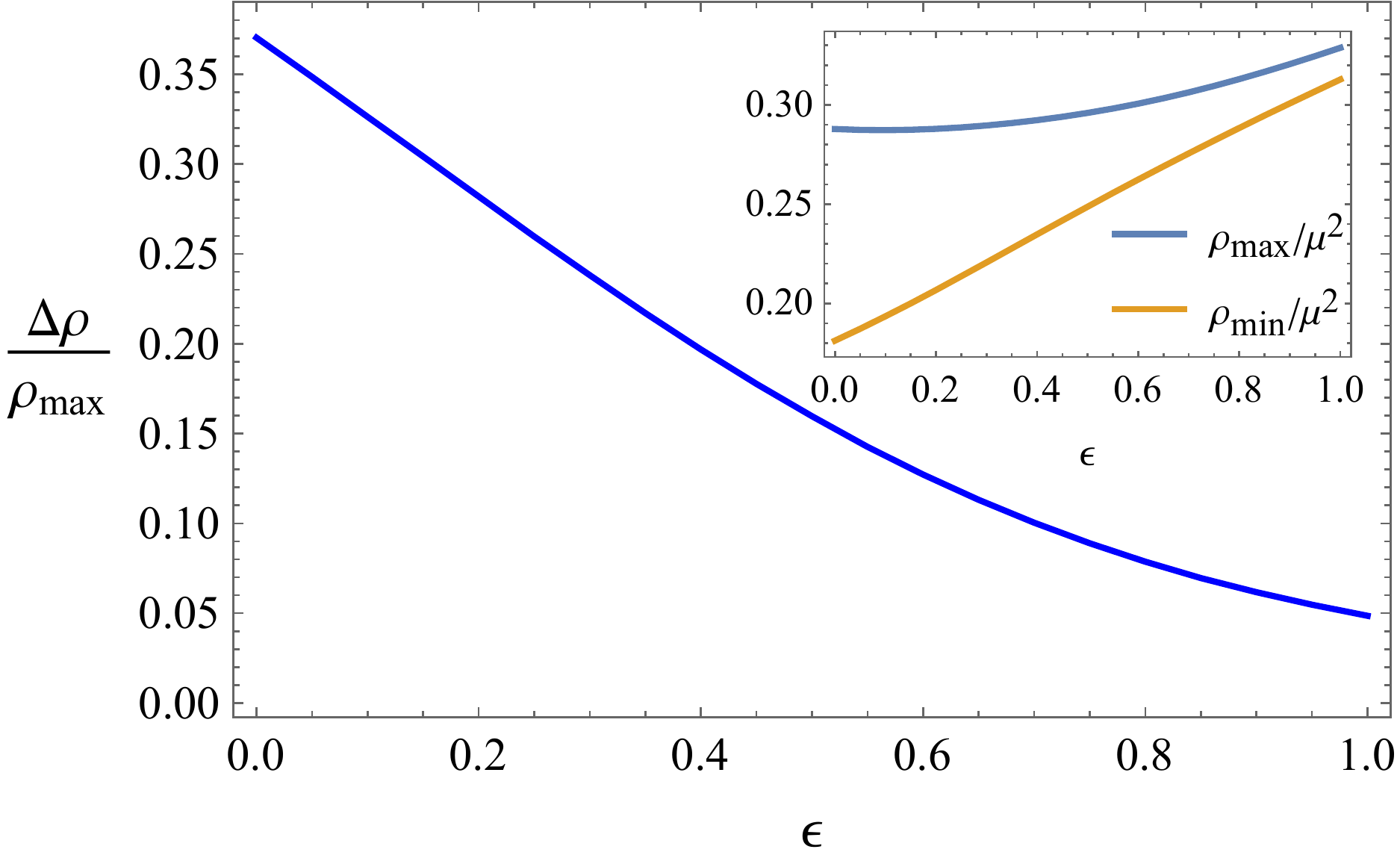}}
\hfill{}\subfloat[Alternative (BEC) case]{\includegraphics[width=0.43\columnwidth]{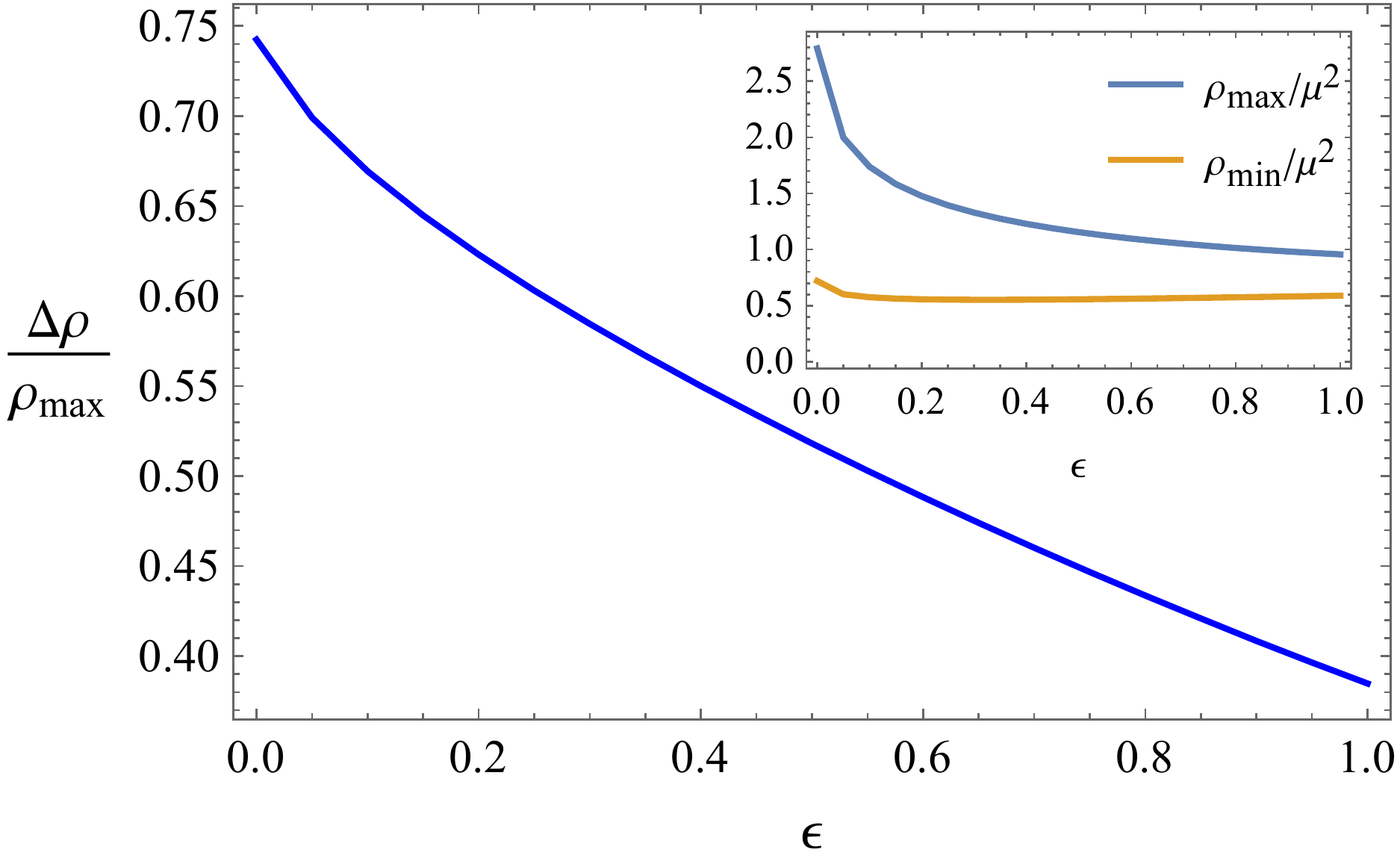}}\hfill{}
\caption{\label{fig:depletion} The depletion of particle number density  as a function of $\epsilon$ at $T/T_{c}=0.5$. The inset panel shows the change in the particle number density far away from the core of the soliton and at the core. In both cases, increasing backreaction reduces the charge depletion.}
\end{figure}

From the asymptotic form of the matter fields, we then read off the expectation value of the charged condensate and the particle number density in the dual field theory. These are shown in  Fig.~\ref{fig:condensate and density}.  As found in Refs.~\cite{Esko2010,Esko2011} in the probe limit for the  BEC  superfluid, the soliton shows a larger depletion fraction which is, however, smaller than $100\%$.\footnote{One possible explanation for this finding is that the temperature here is not low enough. In fact, the solution with backreaction at lower temperature is still very unreliable and difficult to obtain in numerics.} We expect that the depletion will be close to $100\%$ at lower temperature, where backreaction has to be included. The dependence of the depletion factor on temperature for different $\epsilon$ is plotted in Fig.~\ref{fig:depletionvsT}. Contrary to expectations, for the BEC soliton, the depletion in the core is considerably smaller than $100\%$, and even lower than {that} in the probe limit at low temperature. As Fig.~\ref{fig:depletion}  shows, the depletion decreases as the backreaction increases. This behavior is observed in both the BCS and BEC regime of our holographic model. We give a qualitative interpretation of this behavior in terms of the balance between uncondensed and condensed charge in the boundary theory as the homogeneous and normal state is reached under the limit of large backreaction in Appendix \ref{sec:appendixB}. We think that the underlying reason for this behavior is the nature of the condensate zero-temperature IR fixed point, which may be uncharged. In the near future we plan to analyze this fixed point using analytic methods along the lines of Ref.~\cite{Charmousis2010}, and also construct other fixed points which show increasing depletion with decreasing temperature.

\section{A Simple Mechanism for the Snake Instability\label{sec:thermodynamics}}

In this section we turn to the discussion of the thermodynamics of our holographic dark soliton solution, as well as its instabilities. Since the boundary
chemical potential and temperature are fixed, our system is in
the grand canonical ensemble characterized by the grand potential
\begin{equation}
	\text{\ensuremath{\Omega=E-TS-\mu N}}\label{eq:grand potential}\,.
\end{equation}
Here $N$ is the total particle number obtained by integrating the charge density $\rho$ over space.
The internal energy is found to be
\begin{equation}
	E=\int_{\Sigma_{t}}d^{2}x\sqrt{\eta}\left[T_{\mu\nu}\left(\partial_{t}\right)^{\mu}\right]t^{\nu}\label{eq:energy}.
\end{equation} 
Here  $\eta_{\mu\nu}$ 
is the induced metric on the surface $\Sigma_{t}$ at $z=0$ and $t=$ const, 
with unit normal vector $t^{\nu}$, and $T_{\mu\nu}$ is the holographic stress-energy tensor, see Appendix \ref{stressenergytensor} for a detailed calculation. $S$ is the usual black hole entropy, given by 
\begin{equation}
	S=\frac{A_{h}}{4G}=\frac{4\pi z_{h}^{2}}{\epsilon}\int\sqrt{Q_{4}\left(0,x\right)Q_{5}\left(0,x\right)}d^2x.
\end{equation}
\begin{figure}
\centering
\hfill{}\subfloat[Total energy density]{\includegraphics[width=0.45\columnwidth]{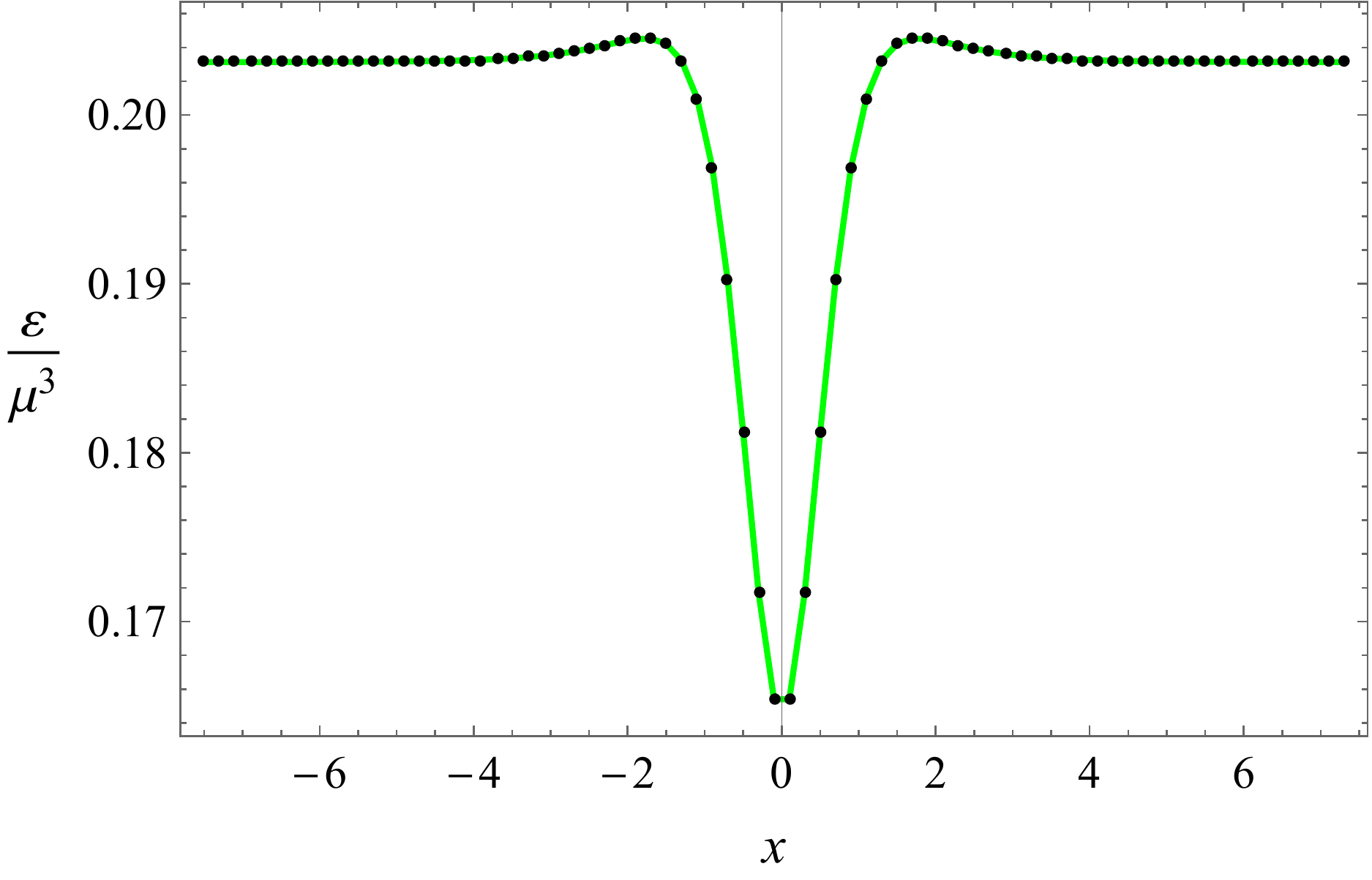}
}\hfill{}\subfloat[Subtracted energy density \label{fig:subtracted energy density}]{\includegraphics[width=0.45\columnwidth]{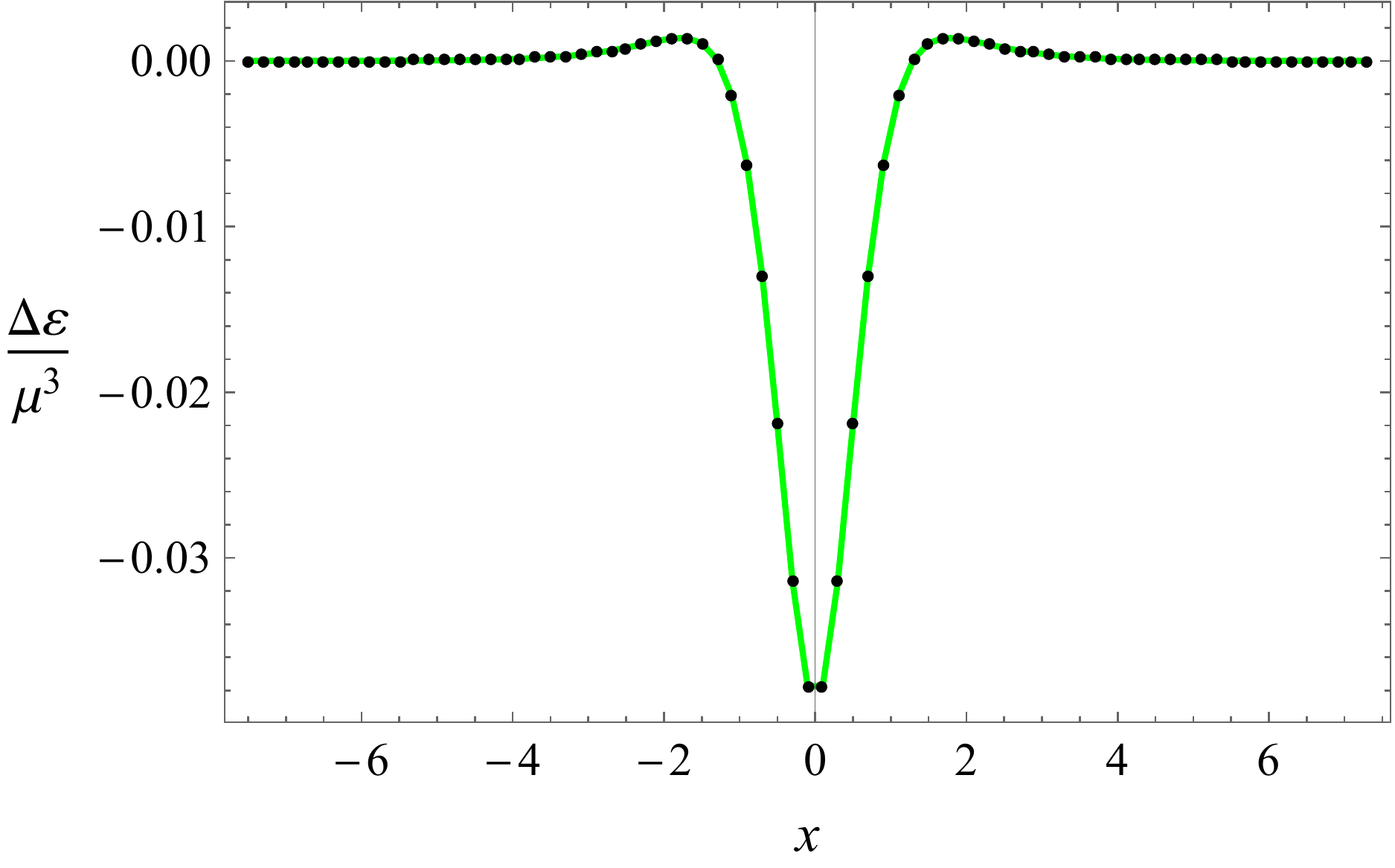}}\hfill{}
\caption{Total (a) and subtracted (b) energy density at $\epsilon=0.25$, $T=0.5T_{c}$. The subtracted energy density, which can be interpreted as the effective mass density of the soliton in the condensate phase, is found to be negative.}
\end{figure}
Since the soliton extends in a noncompact spatial direction, in what follows we will consider densities of the above thermodynamic quantities. % satisfy the Gibbs-Duhem relation 
%\begin{equation}
%\omega=\varepsilon-Ts-\mu\rho,
%\end{equation}
In particular, the grand potential $\Omega$  and the energy $E$ are replaced by their respective densities $\omega$ and  $\varepsilon$. 
Far away from the soliton center, these local quantities will approach
their homogeneous equilibrium values. Therefore,  the soliton core is characterized by 
the difference between these local densities and their equilibrium values. 

The  energy density difference is displayed in
Fig.~\ref{fig:subtracted energy density}, where we see an obvious
energy depletion around the soliton core.\footnote{In this section, we only show the explicit results for the soliton in the BCS superfluid, while the results for the BEC  case are similar.} Upon integration along $x$, this depletion yields a negative effective energy difference $\Delta E = -7.220$ (in units of the chemical potential $\mu$). This energy difference $\Delta E$ can be seen to set the effective mass of the soliton, which is negative as expected for a dark soliton \cite{Fedichev1999}.

\begin{figure}
\centering
\hfill{}\subfloat[Total grand potential density]{\includegraphics[width=0.45\columnwidth]{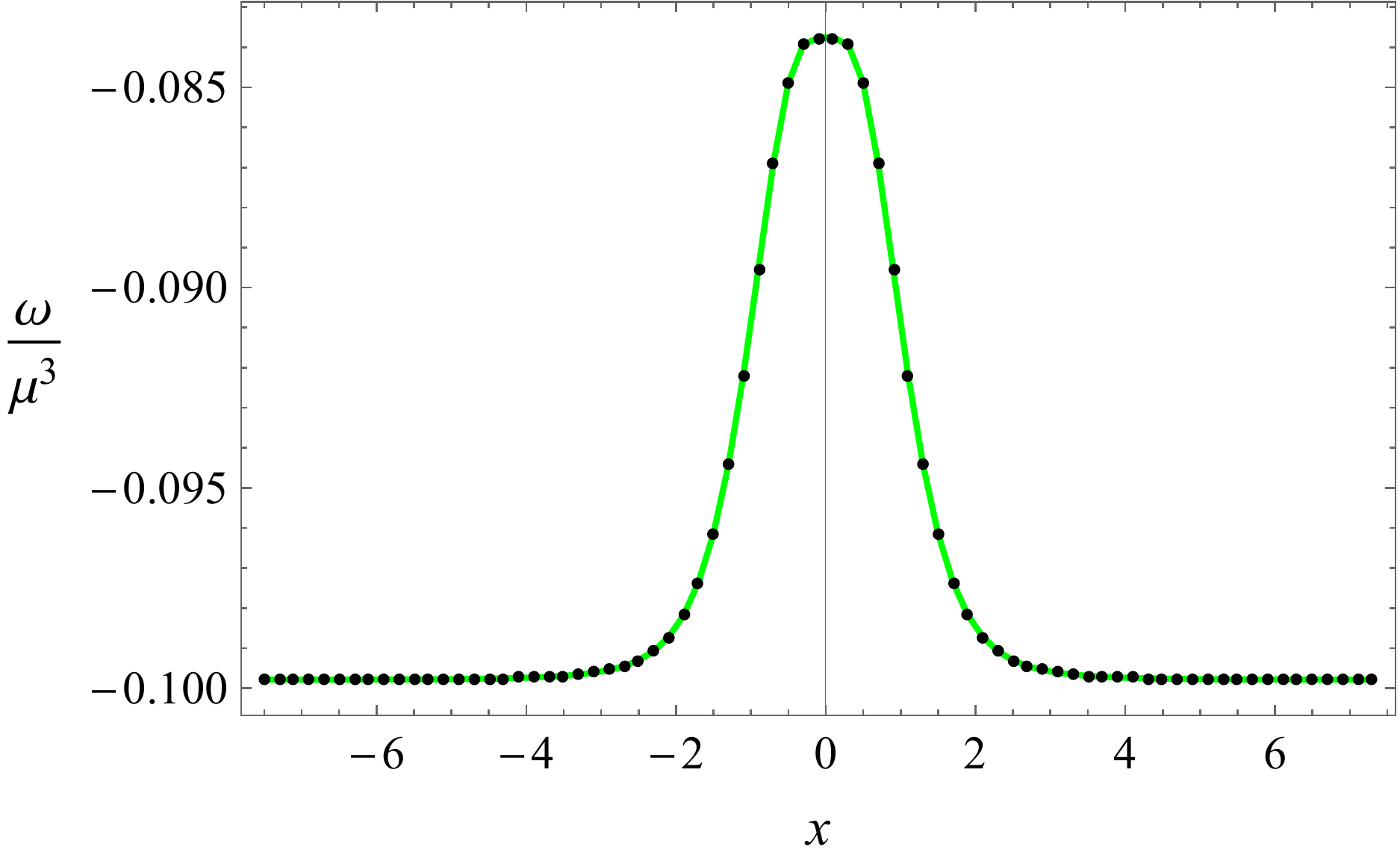}}
\hfill{}\subfloat[Subtracted grand potential density \label{fig:subtracted grand potential density}]{\includegraphics[width=0.45\columnwidth]{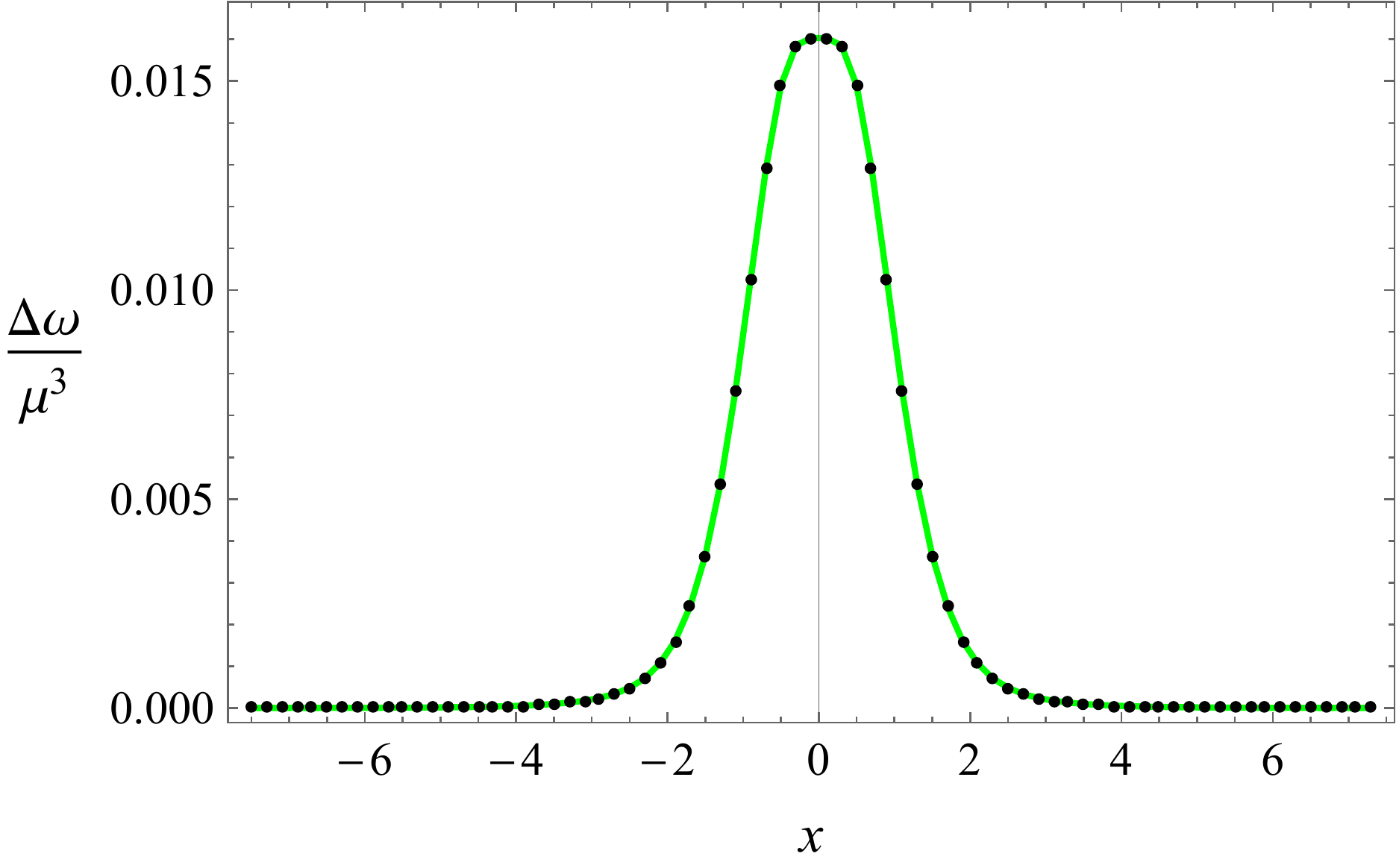}}\hfill{}
\caption{Total (a) and subtracted (b) grand potential density at $\epsilon=0.25$, $T=0.5T_{c}$. After integration over the width of the soliton, the latter becomes the surface tension of the soliton, which is found to be positive.}
\end{figure}

\begin{figure}
\centering
\hfill{}\subfloat[Pressure anisotropy\label{fig:pressure anisotropy}]{\includegraphics[width=0.45\columnwidth]{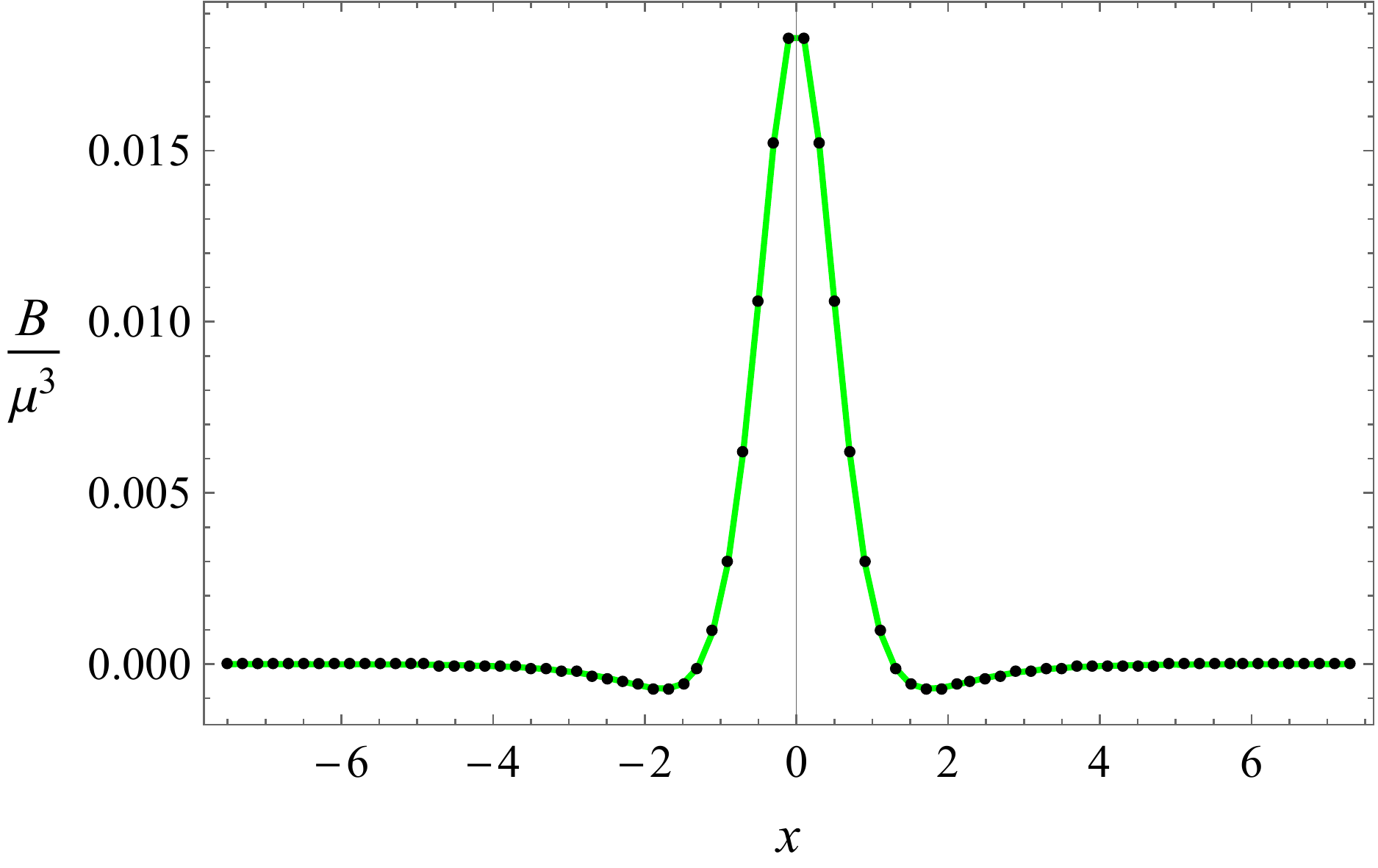}}
\hfill{}\subfloat[Validity of the thermodynamic relation $\omega=-p$]{\includegraphics[width=0.45\columnwidth]{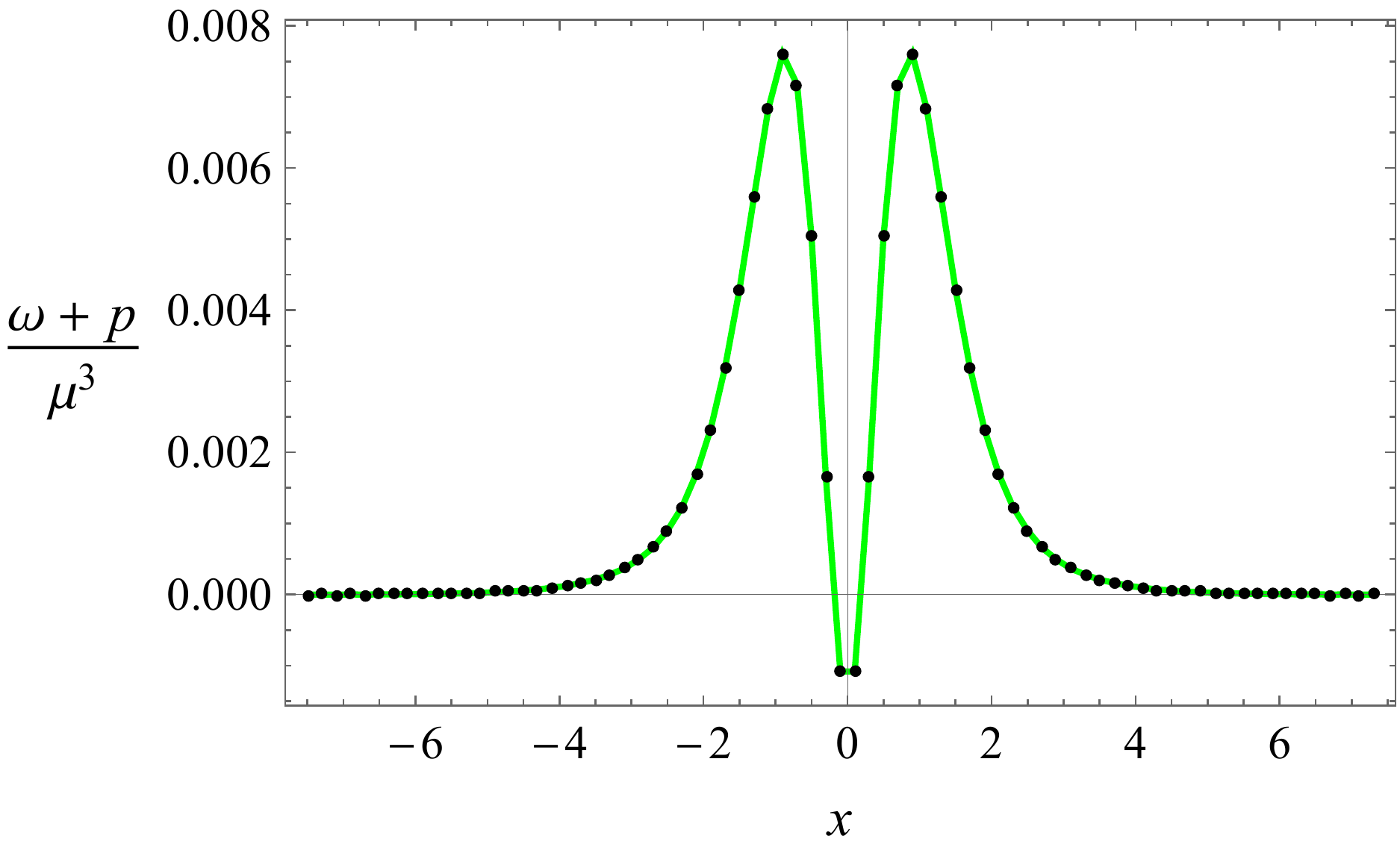}}\hfill{}
\caption{Pressure anisotropy \textbf{(a)} and thermodynamical relation \textbf{(b)} at $\epsilon=0.25$,
$T=0.5T_{c}$. Panel \textbf{(b)} in particular shows that the system is in a thermalized state away from the soliton core.}
\end{figure}

The grand potential density  difference
is plotted in Fig.~\ref{fig:subtracted grand potential density},
from which  we see that there is a grand potential cost for the soliton
with respect to the homogeneous background.  Upon
integration along $x$, the grand potential cost  of the soliton yields the surface tension
coefficient of the soliton. The surface tension coefficient $\sigma$ of a
domain wall such as the soliton is defined 
as the external work $W$ necessary  to enlarge
the surface by a unit area while keeping temperature and chemical potential fixed. 
Under these conditions, the external work is just the increase of
the grand potential due to the enlargement of the domain wall surface,
$W=\Omega-\Omega_{0}$, with $\Omega_{0}$ being the grand potential of the corresponding homogeneous system without the domain wall.
For the case displayed in Fig.~\ref{fig:subtracted grand potential density}, we numerically determine the surface tension in units of the chemical potential to be $\sigma = 6.615$. As a consistency check for our numerics, we also plot the pressure anisotropy $B\equiv p_{x}-p$ in Fig.~\ref{fig:pressure anisotropy}
and check the thermodynamic relation $\omega=-p$, where $p$ is the average pressure, which should hold far away from the soliton center.

With the results for these thermodynamic variables, we can confirm the following explanation for the so-called snake instability of the dark soliton \cite{Cetoli2013}: The soliton moves through the condensate as a heavy, i.e, nonrelativistic, particle \cite{Fedichev1999}. Since it  has a negative effective
mass $M_{eff} = \Delta E<0$, its energy $E_s = \frac{M_eff}{2} \dot q^2$ decreases with increasing velocity $\dot q$. As shown in Refs.~\cite{Fedichev1999}, for a homogeneous solitonic configuration, the velocity grows exponentially. 
This is the so-called  self-acceleration instability \cite{Fedichev1999} of the dark soliton. The self-acceleration originates from the dissipative interaction of the soliton with the surrounding condensate. As discussed in Refs.~\cite{Fedichev1999},  for a homogeneous soliton configuration the self-acceleration terminates once the soliton velocity  reaches the speed of sound, at which point the soliton decays into sound waves that dissipate away in the condensate.

On the other hand, the soliton has a positive surface tension coefficient $\sigma$. As shown in the hydrodynamic approximation in Refs.~\cite{Cetoli2013}, the combined effect of the negative effective mass together with the positive surface tension leads to a growing transverse bending mode with a finite wave vector. The same instability was also found in the holographic quasinormal mode spectrum in Refs.~\cite{Guo2020}. Once this transverse bending mode starts to grow, the self-acceleration instability  enhances the local bending, leading to the formation of a snake-like structure.

\section{Conclusions and Discussions\label{sec:conanddis}}

We investigated the implications of including the gravitational backreaction onto solitons in holographic superfluid systems. We numerically solved Einstein equations coupled with the relevant matter fields. As compared to the probe limit, and contrary to our original expectations,  increasing the backreaction decreases the depletion of the particle number density in the soliton core. We gave a qualitative interpretation of this in terms of the balance of the ratios of condensate and noncondensate over total charge in the dual field theory as a homogeneous state is reached at strong backreaction. Finally, we computed the holographic stress-energy tensor of the system and confirmed a simple holographic explanation for the snake instability of the dark soliton. 

In this work, we restricted ourselves to the asymptotic regimes of BEC and BCS superfluidity. In particular we did not investigate the Robin boundary conditions necessary to model the actual unitary regime, in which the strongly coupled unitary fermion system is expected to live. We plan to investigate the behavior of the soliton in the crossover regime in future work. Such an investigation may in particular provide a better description of the intermediate unitary fermion regime, for which the Bogoliubov-de Gennes theory provides only a broad approximation \cite{Cetoli2013}. Furthermore, our qualitative interpretation of the reduction of the  depletion factor implicitly relies on the assumption of a noncondensate homogeneous infrared fixed point at strong backreaction or low temperatures. In our holographic superfluid strong backreaction implies a smaller critical temperature, and hence the limit of strong backreaction is equivalent to the limit of low temperatures. Only if the homogeneous infrared fixed point is noncondensate, i.e, the condensing operator is irrelevant at the infrared fixed point, can the system return to the noncondensate state in the low-temperature limit. We will investigate the  possible infrared fixed points in our system along the lines of Refs.~\cite{Charmousis2010} to support our qualitative interpretation in future work.  In the thin-domain-wall limit of the soliton, an analytic treatment of the domain wall in terms of a brane with junction conditions is also conceivable.

\begin{acknowledgments}
	We would like to thank Xin Li, Hong Liu, Jie Ren, Ya-Wen Sun, Xiaoning Wu and Hongbao Zhang for useful discussions. In particular,  Zhongshan Xu is grateful to Jie Ren for his explanations concerning the DeTurck method. This work is partially supported by the NSFC with Grant No.~11975235 and 11847229. Y. D. is supported by the China Scholarship Council. Y. D., J. E and R. M acknowledge financial support through the Deutsche Forschungsgemeinschaft (DFG, German Research Foundation), project-id 258499086 - SFB 1170 ``ToCoTronics'', and through the W\"urzburg-Dresden Cluster of Excellence on Complexity and Topology in Quantum Matter - ct.qmat (EXC 2147, project-id 39085490). Y. T. is also supported by the \textquotedblleft Strategic Priority Research Program of the Chinese Academy of Sciences\textquotedblright{} with Grant No.XDB23030000. Z.-Y. X also acknowledges the support from the National Postdoctoral Program for Innovative Talents BX20180318, funded by the China Postdoctoral Science Foundation.
\end{acknowledgments}

\appendix

\section{Holographic stress-energy tensor}\label{stressenergytensor}
In order to compute the holographic stress energy tensor, following
the process of Refs.~\cite{Santos2014,Tallarita2019}, we need
to find the asymptotic expansion of the metric at the conformal boundary. 
The expansion is obtained by solving Eq.~\eqref{eq:Deturck} with boundary conditions
order by order in $\left(1-r\right)$ and in addition imposing $\xi^{\mu}=0$,
\begin{subequations}
	\begin{equation}
		Q_{i}\left(r,x\right)=1-\frac{\epsilon Q_{6}^{\left(0\right)}\left(x\right)}{z_{h}^{2}}\left(1-r\right)^{2}+q_{i}\left(x\right)\left(1-r\right)^{3}+O\left[\left(1-r\right)^{4}\right],i=1,4,5,
	\end{equation}
	\begin{equation}
		Q_{2}\left(r,x\right)=1+\frac{8\epsilon Q_{6}^{\left(0\right)}\left(x\right)Q_{6}^{\left(1\right)}\left(x\right)}{3z_{h}^{2}}\left(1-r\right)^{3}+O\left[\left(1-r\right)^{4}\right],
	\end{equation}
	\begin{equation}
		Q_{3}\left(r,x\right)=\frac{2\epsilon Q_{6}^{\left(0\right)}\left(x\right)\partial_{x}Q_{6}^{\left(0\right)}\left(x\right)}{z_{h}^{3}}\left(1-r\right)^{3}+O\left[\left(1-r\right)^{4}\right],
	\end{equation}
	\begin{equation}
		Q_{6}\left(r,x\right)=Q_{6}^{\left(0\right)}\left(x\right)+Q_{6}^{\left(1\right)}\left(x\right)\left(1-r\right)+\cdots,
	\end{equation}
	\begin{equation}
		Q_{7}\left(r,x\right)=Q_{7}^{\left(0\right)}\left(x\right)+Q_{7}^{\left(1\right)}\left(x\right)\left(1-r\right)+\cdots.
	\end{equation}
\end{subequations}
Here $q_{1}\left(x\right),q_{4}\left(x\right),q_{5}\left(x\right)$
satisfy Eqs.~\eqref{eq:traceless and conserved} related to the tracelessness 
and conservation of the boundary stress-energy tensor,
\begin{equation}
	T_{i}^{i}=0,\ \ \ \partial_{i}T^{ij}=0.
\end{equation}
Using these relations, one can explicitly show the following conditions:
\begin{subequations}
	\label{eq:traceless and conserved}
	
	\begin{equation}
		q_{1}\left(x\right)+q_{4}\left(x\right)+q_{5}\left(x\right)=-\frac{\epsilon Q_{6}^{\left(0\right)}\left(x\right)\left(-3Q_{6}^{\left(0\right)}\left(x\right)+8Q_{6}^{\left(1\right)}\left(x\right)\right)}{z_{h}^{2}},
	\end{equation}
	
	\begin{equation}
		\partial_{x}q_{4}\left(r,x\right)=-\frac{2\epsilon\left[\left(-3Q_{6}^{\left(0\right)}\left(x\right)+8Q_{6}^{\left(1\right)}\left(x\right)\right)\partial_{x}Q_{6}^{\left(0\right)}\left(x\right)+2Q_{6}^{\left(0\right)}\left(x\right)\partial_{x}Q_{6}^{\left(1\right)}\left(x\right)\right]}{3z_{h}^{2}}.
	\end{equation}
\end{subequations}

Having obtained the asymptotic behavior of the metric functions, one then changes  to Fefferman-Graham coordinates $\left(z,v\right)$ by an expansion
of the series \eqref{eq:coodtran} and demands $g_{zz}=\frac{1}{z^{2}}$
and $g_{zv}=0$ to determine the two functions $\left\{ a_{k}\left(v\right),b_{k}\left(v\right)\right\} $
order by order in $z$. Here we provide the first few terms necessary for the
computation of holographic stress-energy tensor:
\begin{subequations}
	\begin{equation}
		\begin{cases}
			r=1-\frac{z_{h}}{2}z+\sum_{k=2}^{\infty}a_{k}\left(v\right)z^{k},\\
			x=v+\sum_{k=1}^{\infty}b_{k}\left(v\right)z^{k},
		\end{cases}\label{eq:coodtran}
	\end{equation}
	\begin{equation}
		a_{2}\left(v\right)=-\frac{z_{h}^{2}}{8},\ \ \ a_{3}\left(v\right)=-\frac{z_{h}^{3}}{16},
	\end{equation}
	\begin{equation}
		a_{4}\left(v\right)=\frac{z_{h}^{2}\left(24\epsilon\mu^{2}+51z_{h}^{2}+32\epsilon Q_{6}^{\left(0\right)}\left(v\right)Q_{6}^{\left(1\right)}\left(v\right)\right)}{1152},
	\end{equation}
	\begin{equation}
		b_{1}\left(v\right)=b_{2}\left(v\right)=b_{3}\left(v\right)=0,
	\end{equation}
	\begin{equation}
		b_{4}\left(v\right)=-\frac{\epsilon Q_{6}^{\left(0\right)}\left(v\right)\partial_{v}Q_{6}^{\left(0\right)}\left(v\right)}{16}.
	\end{equation}
\end{subequations}

Finally, the holographic stress-energy tensor is computed by using Eq.~\eqref{eq:stressstand}
in the standard quantization and Eq.~\eqref{eq:stressalter} in the alternative quantization \cite{Vijay1999,Skenderis2001},
\begin{subequations}
	\label{eq:holostresstensor}
	
	\begin{equation}
		T_{\mu\nu}=\frac{1}{\kappa_{4}^{2}}\lim_{z\rightarrow0}\frac{1}{z}\left(K_{\mu\nu}-\gamma_{\mu\nu}K-2\gamma_{\mu\nu}-\frac{\epsilon}{2}\left|\Psi\right|^{2}\gamma_{\mu\nu}\right),\label{eq:stressstand}
	\end{equation}
	\begin{equation}
		T_{\mu\nu}=\frac{1}{\kappa_{4}^{2}}\lim_{z\rightarrow0}\frac{1}{z}\left[K_{\mu\nu}-\gamma_{\mu\nu}K-2\gamma_{\mu\nu}+\frac{\epsilon}{2}\left(-\Psi^{\dagger}n^{\sigma}D_{\sigma}\Psi-C.C.+\left|\Psi\right|^{2}\right)\gamma_{\mu\nu}\right].\label{eq:stressalter}
	\end{equation}
\end{subequations}
Here $K_{\mu\nu}$ is the extrinsic curvature associated with an inward-pointing 
unit normal vector $n^{\sigma}$ on the constant $z=\epsilon$ surface near 
the boundary. $\gamma_{\mu\nu}$ is the induced metric on the cutoff
surface. The last term in Eq.~\eqref{eq:holostresstensor} cancels the
divergences due to the presence of the scalar field \cite{Guo2016}.

\section{Particle number (charge) change with backreaction}\label{sec:appendixB}
\begin{figure}
\centering
\hfill{}\subfloat[Total particle number density.]{\includegraphics[scale=0.3]{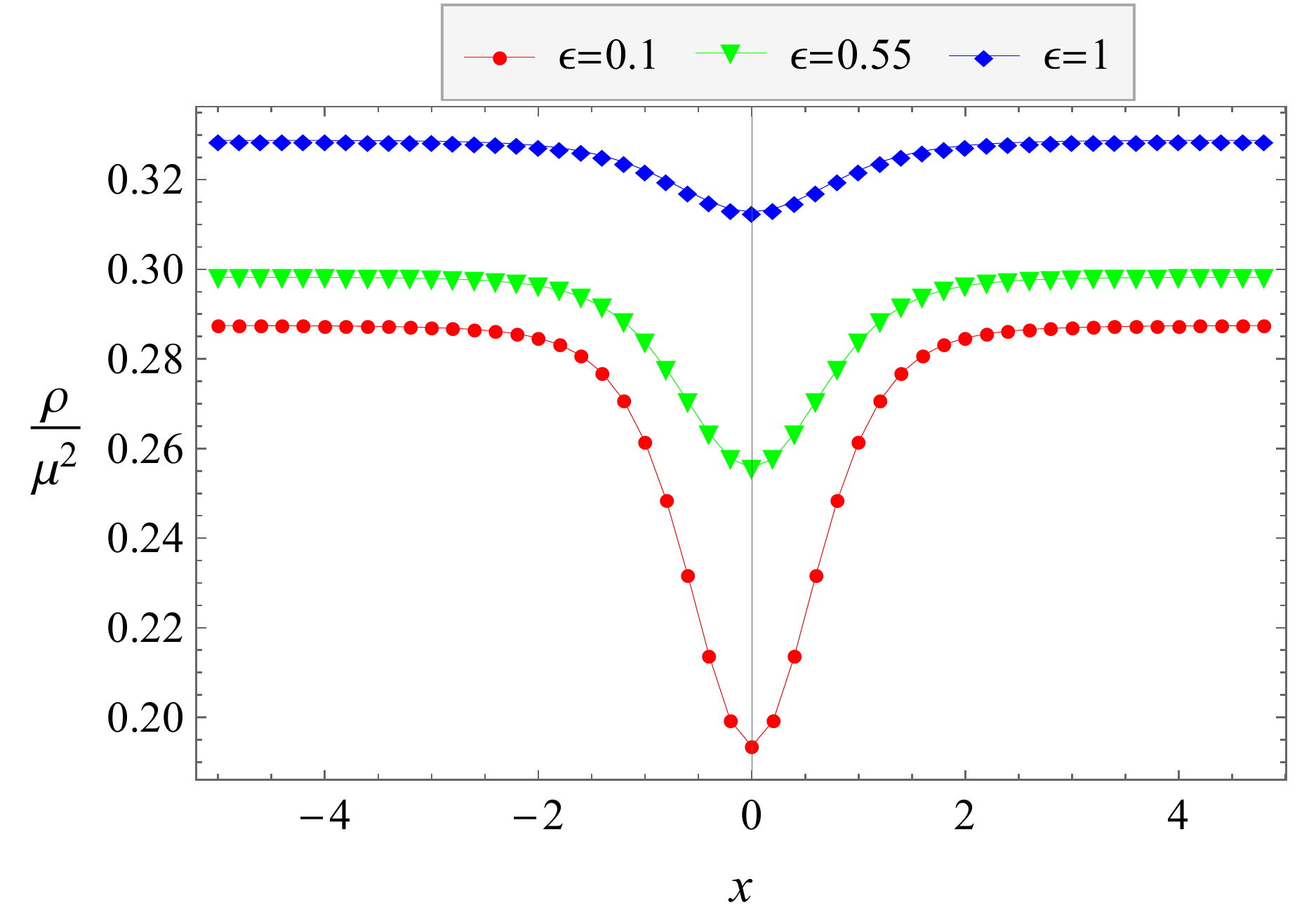}}
\hfill{}\subfloat[Condensed particle number density.]{\includegraphics[scale=0.3]{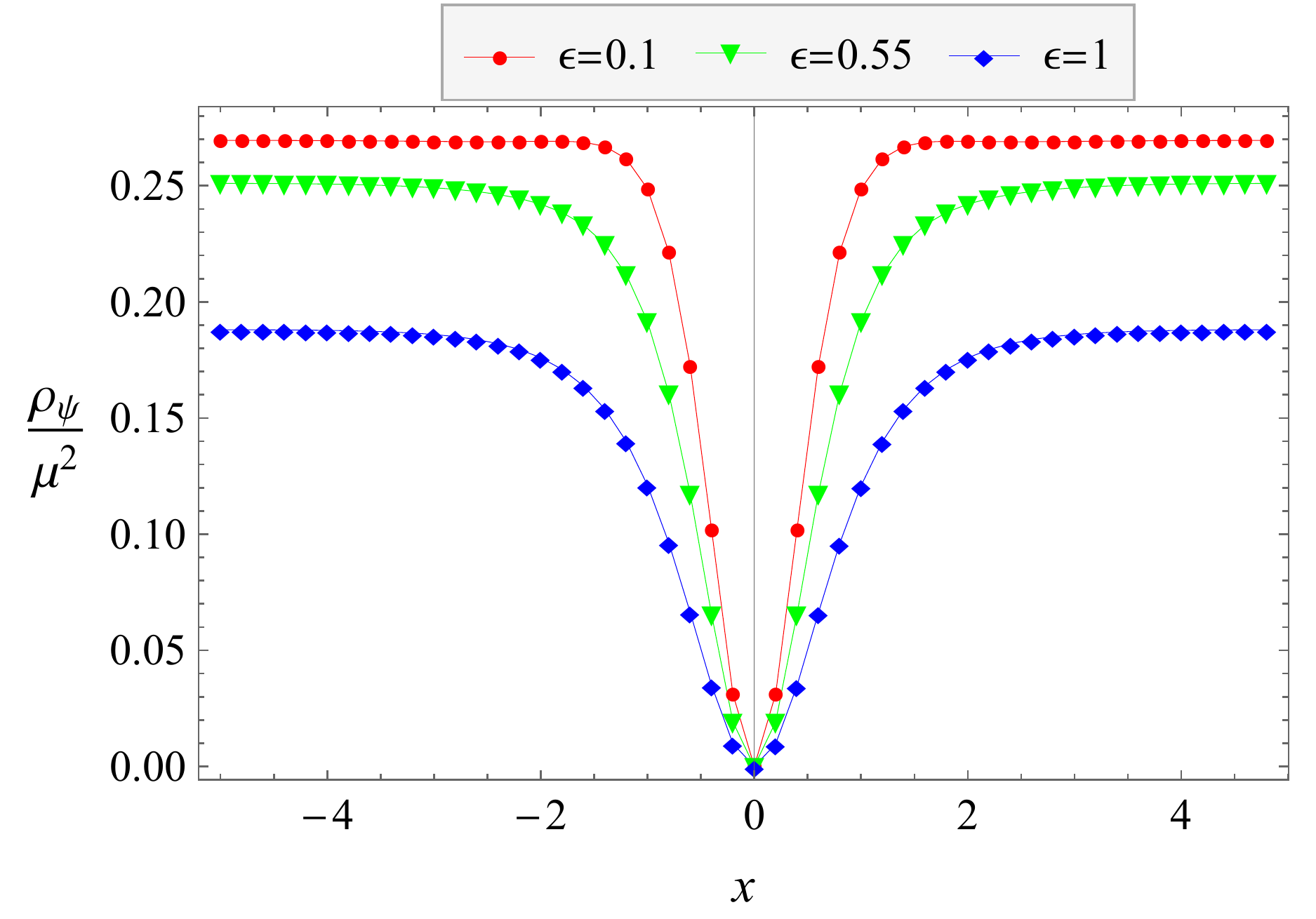}} 
\hfill{}

\hfill{}\subfloat[Uncondensed particle number density.]{\includegraphics[scale=0.3]{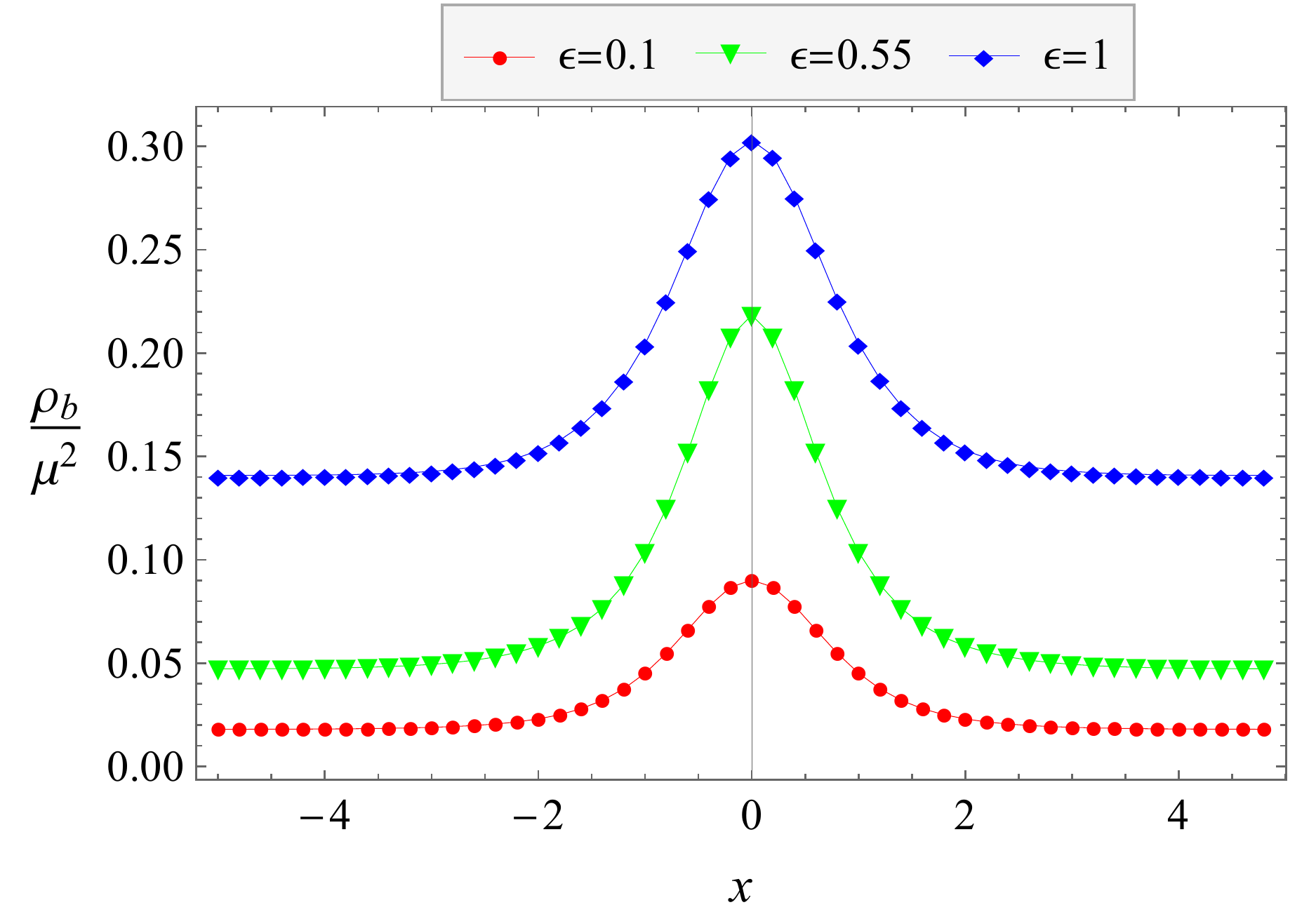}} \hfill{}\subfloat[Soliton contribution.\label{fig:fluxO2}]{\includegraphics[scale=0.3]{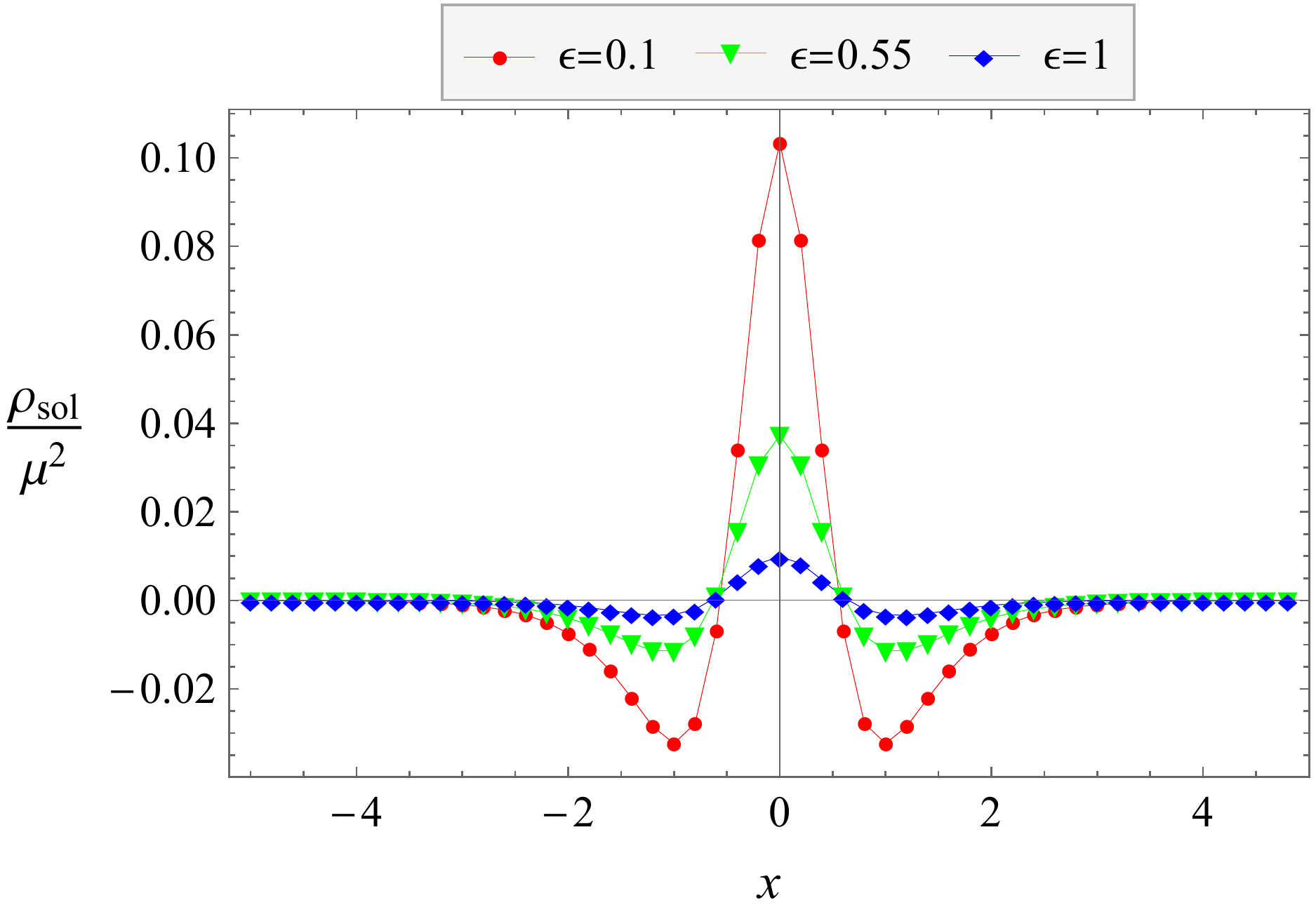}}
\hfill{}
\caption{\label{fig:chargedensityO2}Change of the total charge density (a), condensed charge density (b), uncondensed charge density (c) and soliton contribution (d) with backreaction at $T=0.5T_c$ in the standard (BCS) case.}
\end{figure}
Here we give a qualitative interpretation of the depletion that decreases with increasing backreaction, in terms of the balance between uncondensed and condensed charge in the boundary theory as the homogeneous and normal state is reached under the limit of large backreaction. For our static but inhomogeneous charged scalar and gauge field configuration, by integrating  the $t$ component of the Maxwell equations \eqref{eq:Maxwell} over the holographic $r$ coordinate, one obtains the total particle number density in terms of three contributions,
\begin{equation}
	\sqrt{-g}F^{tr}|_{r=1}=\sqrt{-g}F^{tr}|_{r=0}+\int_{0}^{1}(-\sqrt{-g}J^t)dr+\int_{0}^{1}\partial_{x}\sqrt{-g}F^{xt}dr,\label{eq:numdensity}
\end{equation}

\begin{equation}
	J^{\nu}=ig^{\mu\nu}\left[\Psi^{\dagger}\left(D_{\mu}\Psi\right)-\Psi\left(D_{\mu}\Psi\right)^{\dagger}\right].
\end{equation}
On the right-hand side of Eq.~\eqref{eq:numdensity}, the first term yields the uncondensed particle number density, given by the electric flux evaluated at the horizon. The second term is the condensed particle number density. The third term is a contribution arising from  the inhomogeneous configuration of the soliton.
\begin{equation}
	\underbrace{\int d^2x\sqrt{-g}F^{tr}|_{r=1}}\limits_{N}=\underbrace{\int d^2x\sqrt{-g}F^{tr}|_{r=0}}\limits_{N_{b}}+\underbrace{\int d^2x\int_{0}^{1}(-\sqrt{-g}J^t)dr}\limits_{N_{\psi}}. \label{eq:charge}
\end{equation}

\begin{figure}
\centering
\hfill{}\subfloat[Total particle number density.]{\includegraphics[scale=0.3]{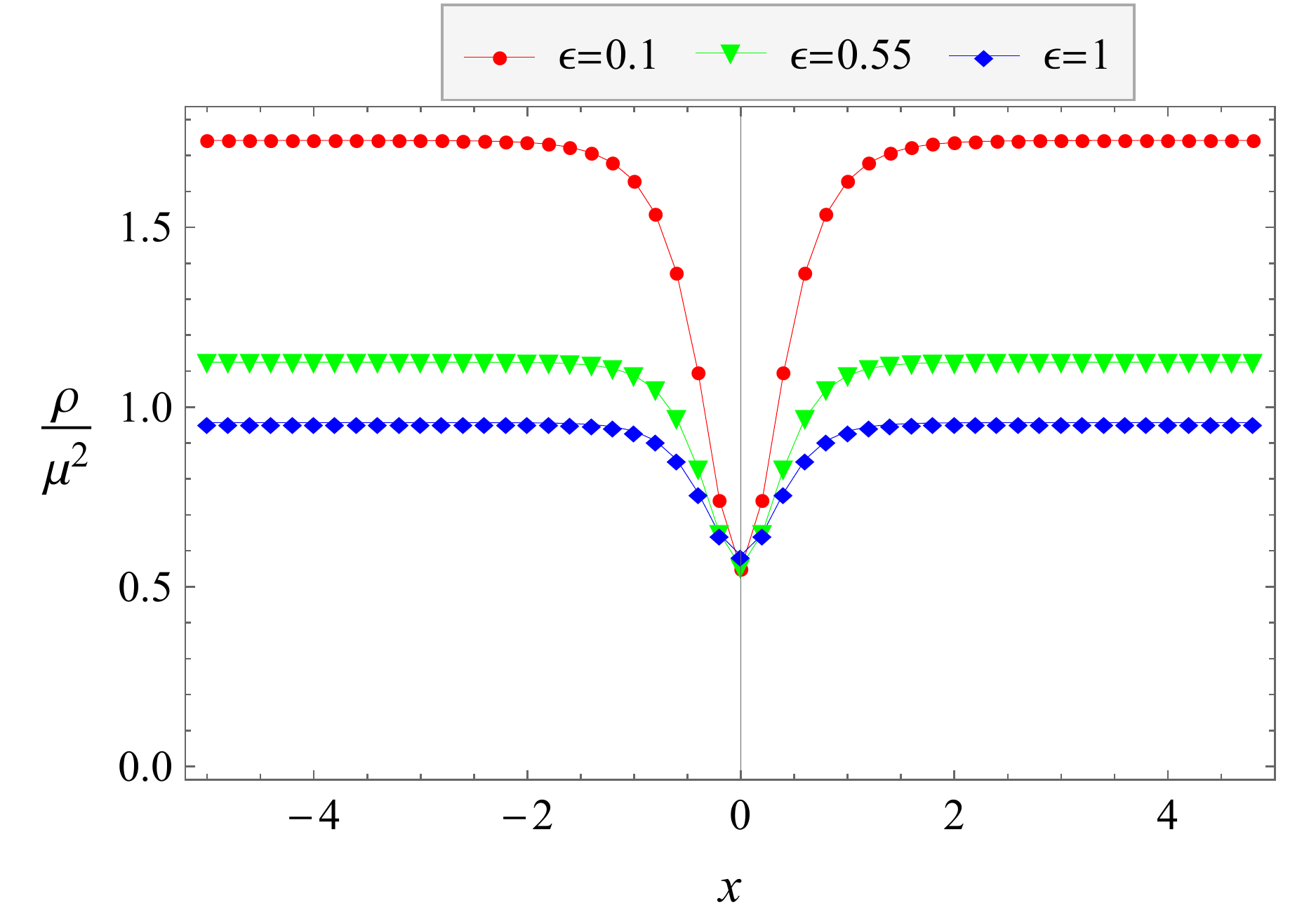}}
\hfill{}\subfloat[Condensed particle number density.]{\includegraphics[scale=0.3]{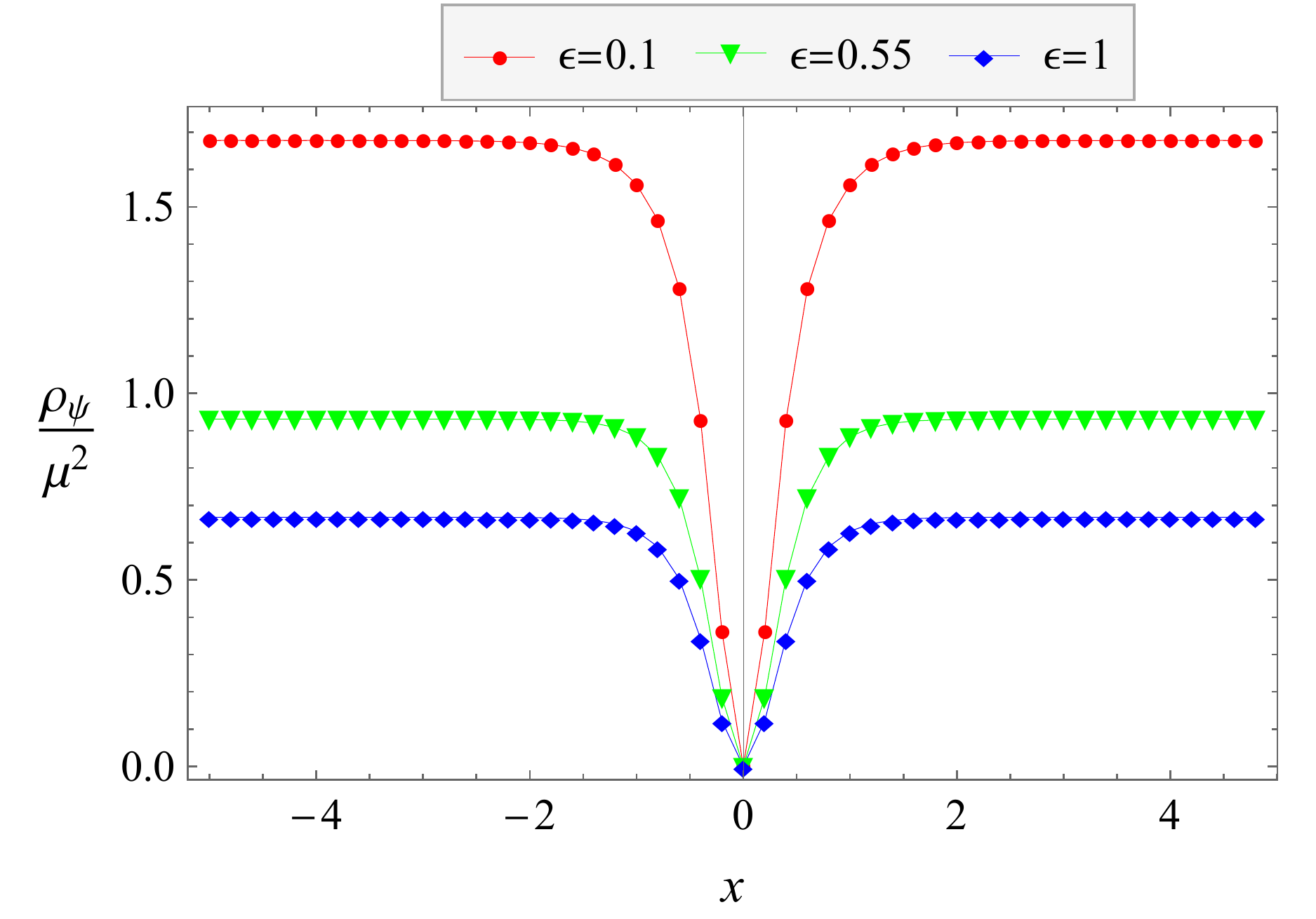}} 
\hfill{}

\hfill{}\subfloat[Uncondensed particle number density.]{\includegraphics[scale=0.3]{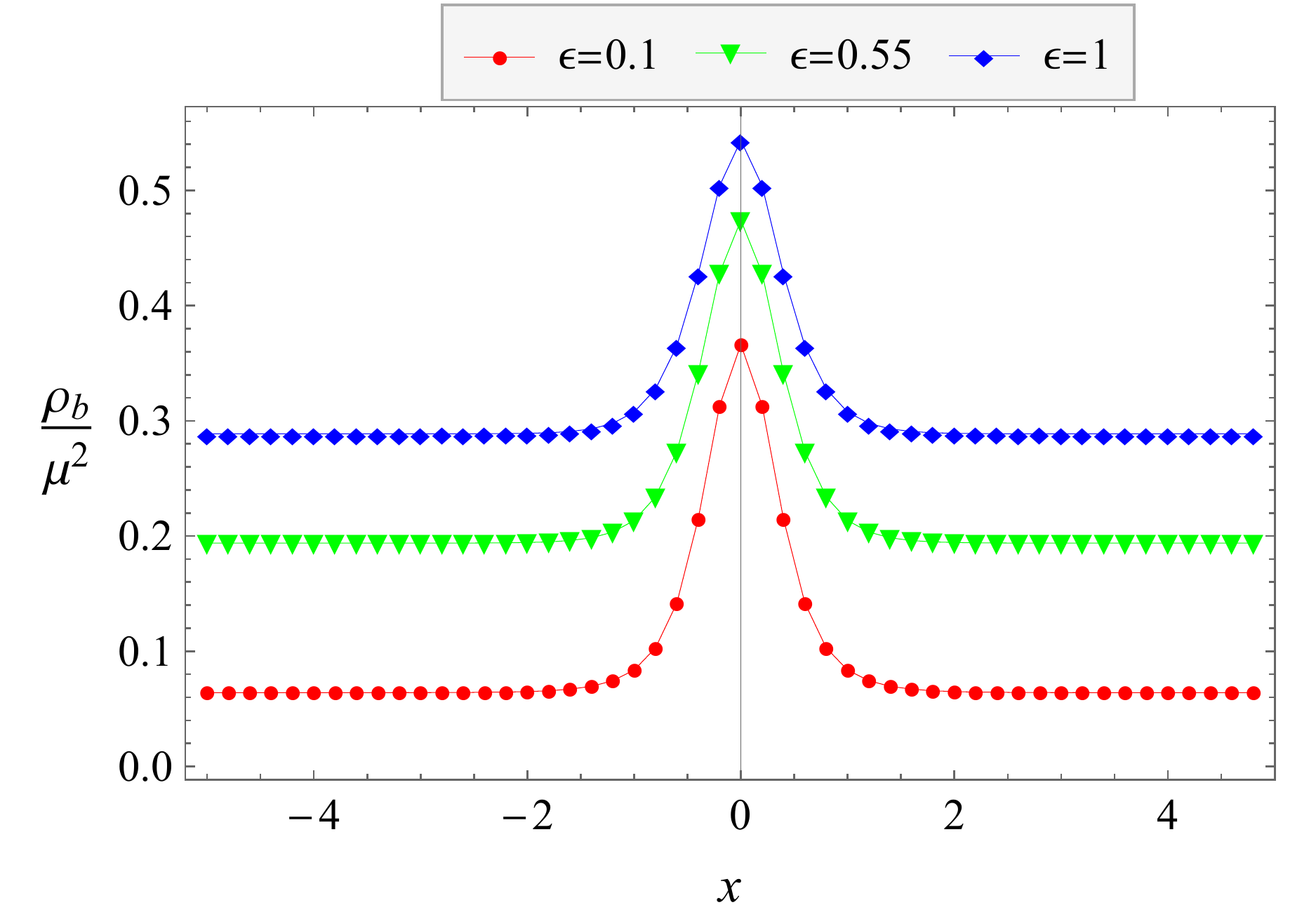}} \hfill{}\subfloat[Soliton contribution.\label{fig:fluxO1}]{\includegraphics[scale=0.3]{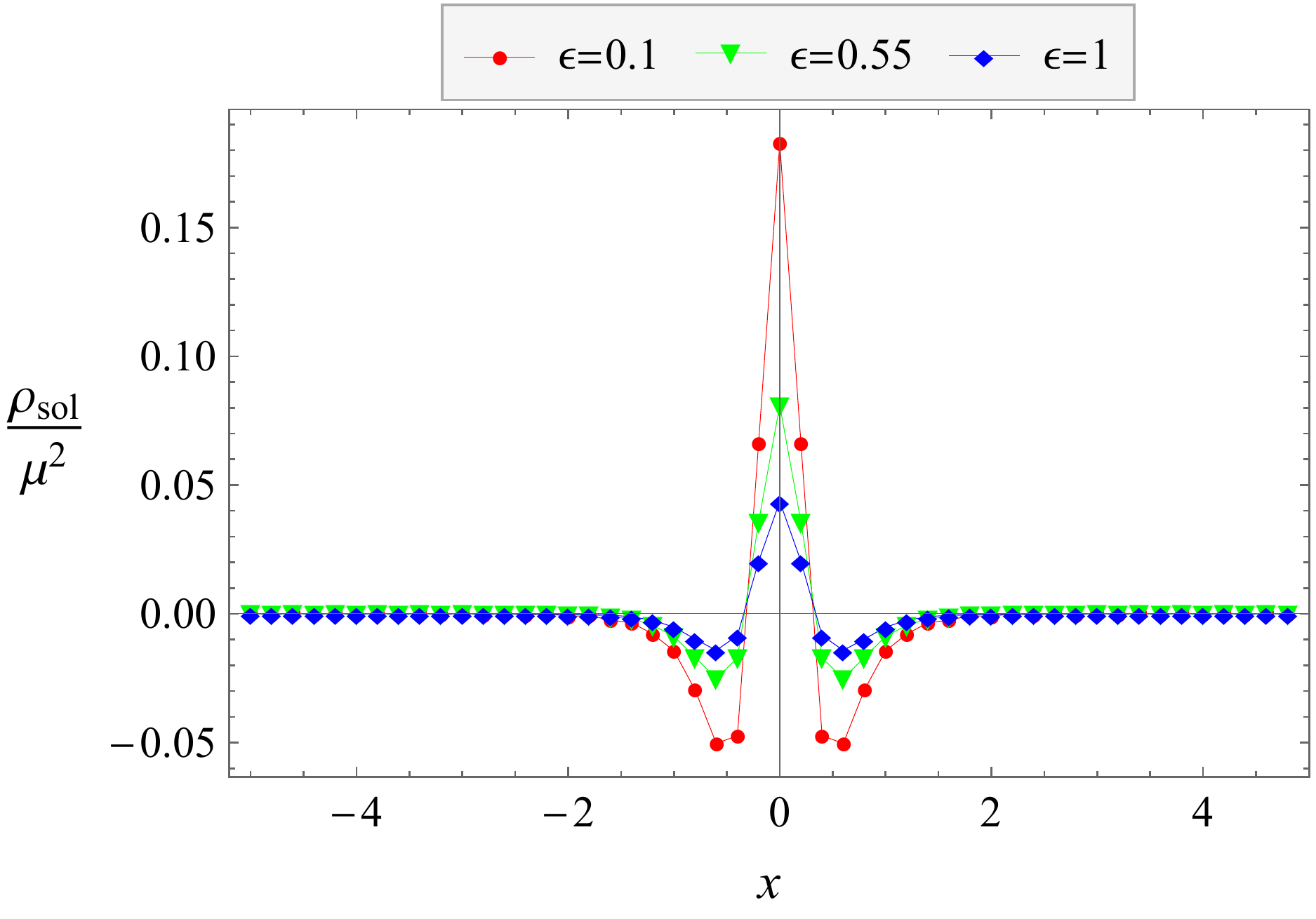}}
\hfill{}
\caption{\label{fig:chargedensityO1}Change of the total charge density (a), condensed charge density (b), uncondensed charge density (c) and soliton contribution (d) with backreaction at $T=0.5T_c$ in the alternative (BEC) case.}
\end{figure}
\begin{figure}
\centering
\hfill{}\subfloat[Standard (BCS) case\label{fig:Otoea}]{\includegraphics[width=0.43\columnwidth]{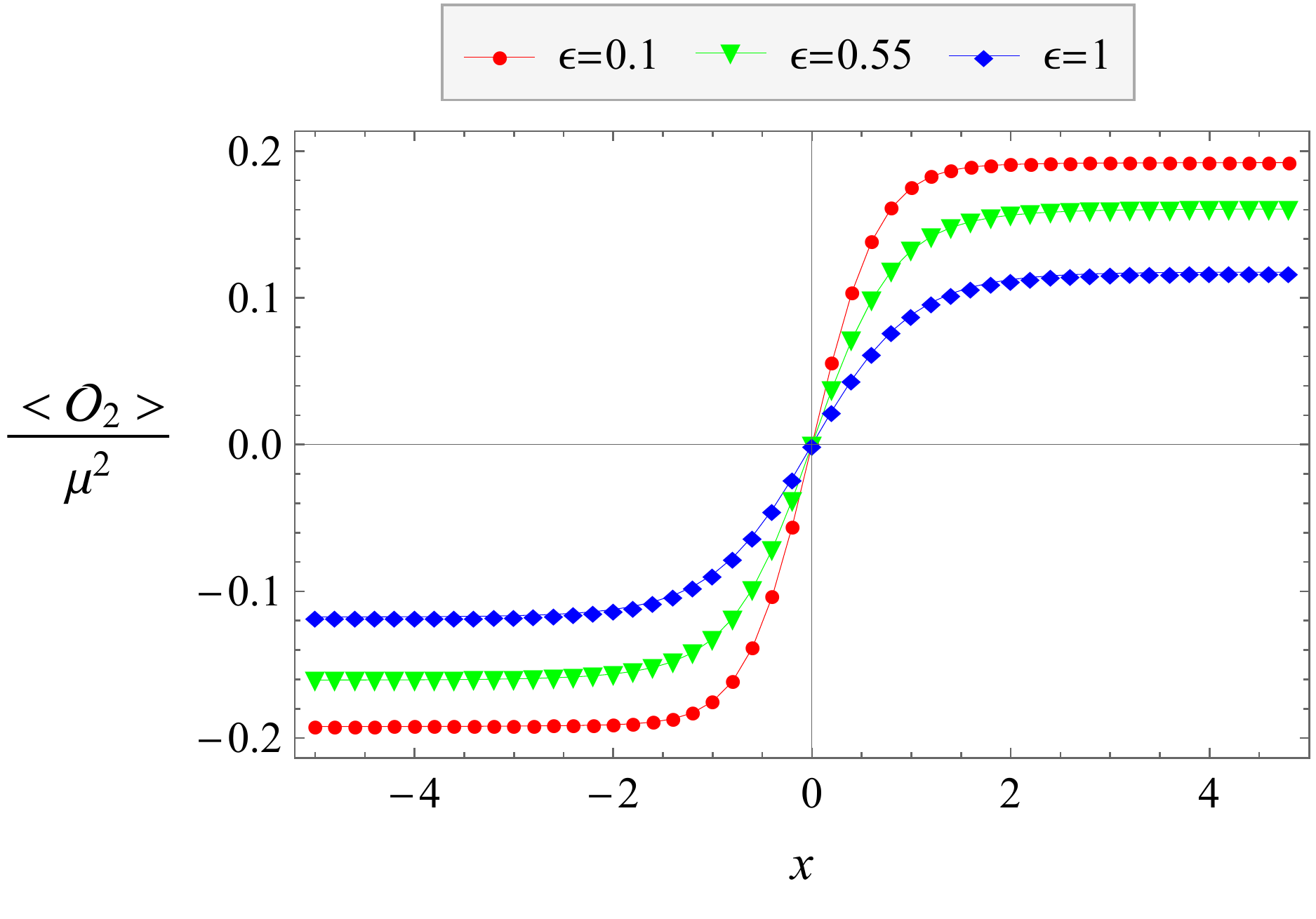}}
\hfill{}\subfloat[Alternative (BEC) case\label{fig:Otoeb}]{\includegraphics[width=0.43\columnwidth]{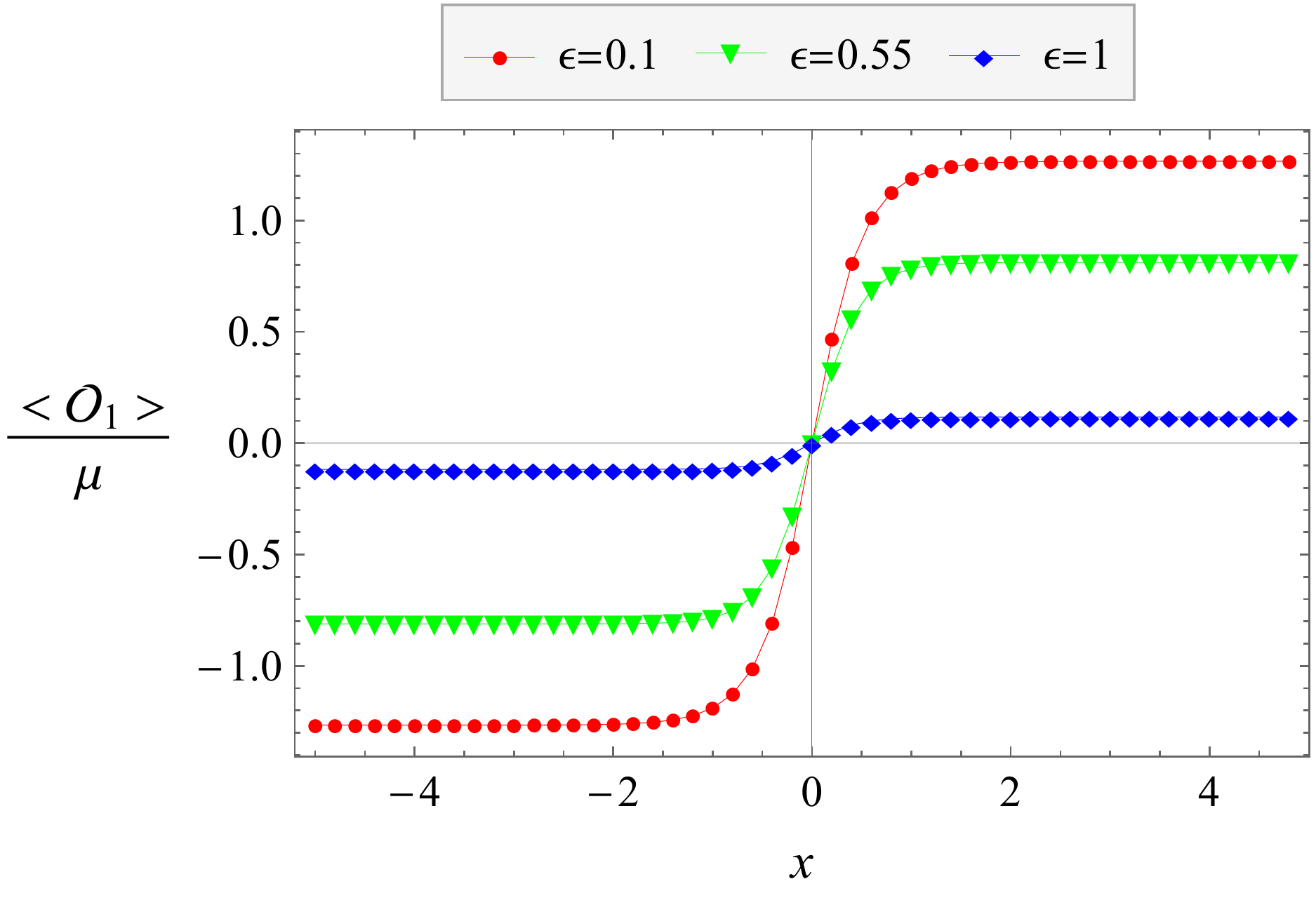}}\hfill{}
\caption{\label{fig:Otoe}The condensate  as a function of $\epsilon$ at $T/T_{c}=0.5$.}
\end{figure}
A relation between the total charge $N$, uncondensed charge $N_b$ and condensed charge $N_{\psi}$ can be obtained by integrating \eqref{eq:numdensity} over the spatial part.\footnote{The spatial domain of all integrations above is $\left(-\frac{L_{x}}{2},\frac{L_{x}}{2}\right)\times\left(-\frac{L_{y}}{2},\frac{L_{y}}{2}\right)$. Due to the  translation symmetry along the $y$ direction, we normalize these charges with regard to $L_{y}$.} After integration, the soliton contribution in Eq.~\eqref{eq:numdensity} becomes a total derivative and hence vanishes if we integrate over a symmetric interval around $x=0$. The remaining terms in Eq.~\eqref{eq:numdensity} yield the relation \eqref{eq:charge} in which we denote the three terms in order as $N$, $N_b$, and $N_{\psi}$.  The ratios of uncondensed and condensed  charge over the total charge then have to add up to one,
\begin{equation}\label{eq:chargeratios}
	\frac{N_b}{N}+\frac{N_{\psi}}{N}=1.
\end{equation}

\begin{figure}
	\centering
	\hfill{}\subfloat[Standard (BCS) case]{\includegraphics[width=0.43\columnwidth]{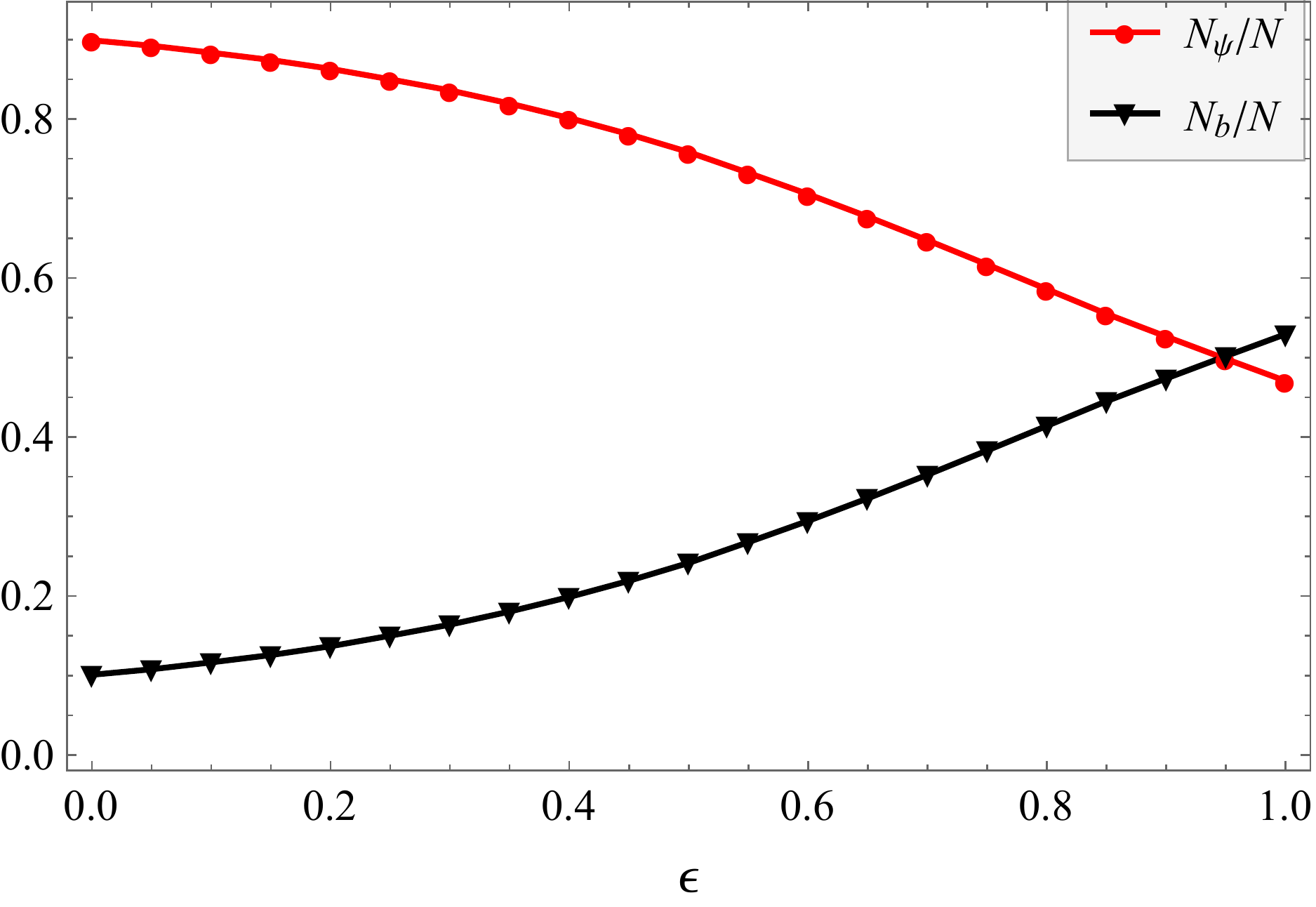}
		
	}\hfill{}\subfloat[Alternative (BEC) case ]{\includegraphics[width=0.43\columnwidth]{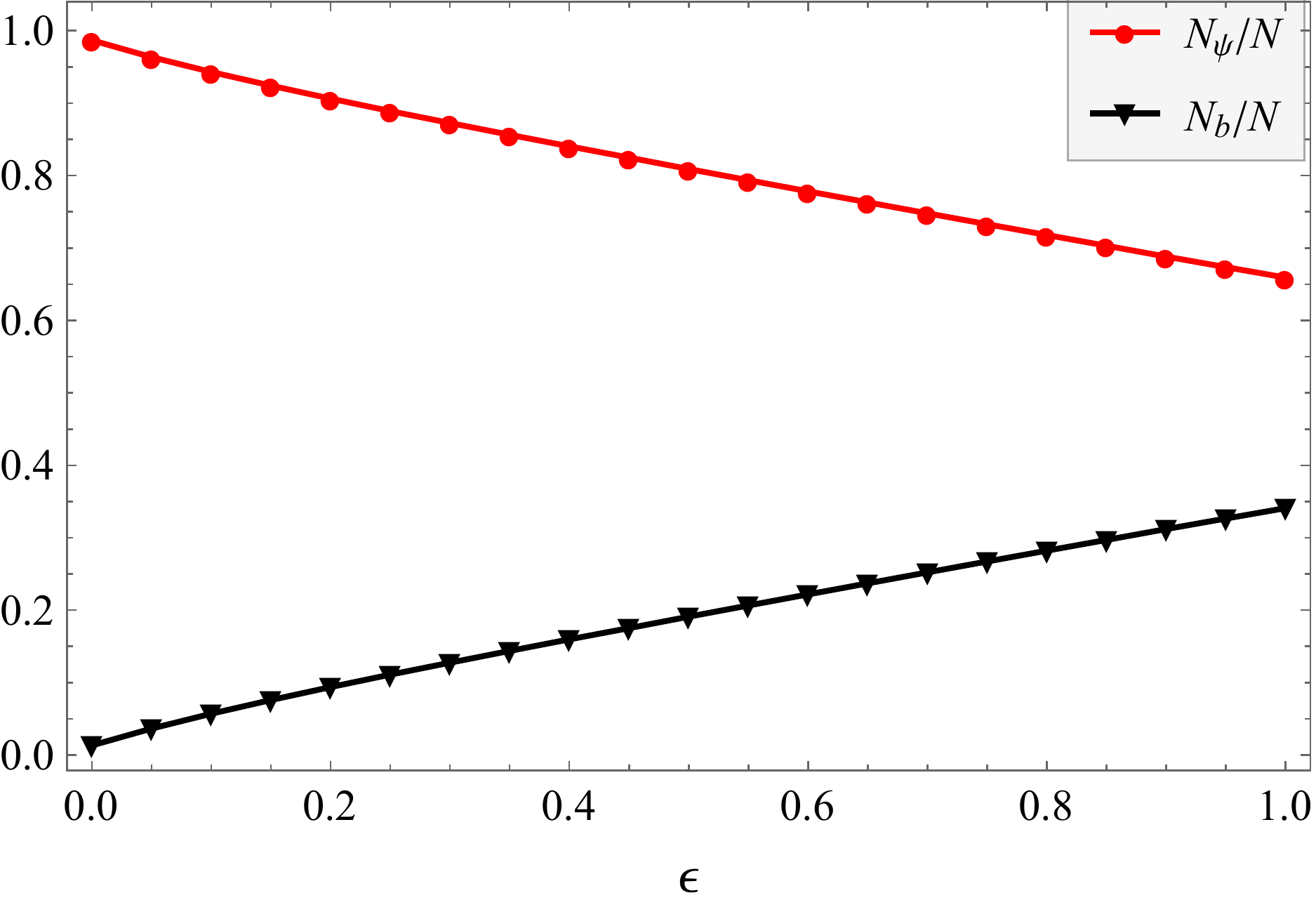}}\hfill{}
	
	\caption{\label{fig:ratio}The ratio of two kinds of particle number  as a function of $\epsilon$ at $T/T_{c}=0.5$.}
\end{figure}

As can be seen from Figs.~\ref{fig:chargedensityO2} and \ref{fig:chargedensityO1}, the backreaction suppresses the condensed particle density and promotes the uncondensed particle number density. On the other hand, from Fig.~\ref{fig:Otoe}, the condensate (order parameter) far away from the soliton core also decreases strongly with increasing backreaction. It seems that the system returns to a homogeneous noncondensate state in the limit of large backreaction. This is, in particular, obvious for the BEC case [Fig.~\ref{fig:Otoeb}], but the trend is also obvious for the BCS case [Fig.~\ref{fig:Otoea}]. Furthermore, this interpretation of the data is  consistent with the general expectation that backreaction inhibits the formation of the condensate in a holographic superconductor \cite{Wang2016}. The suppression of the condensate  implies that the ratio of condensate to total charge monotonically decreases with increasing backreaction. This can be seen from Fig.~\ref{fig:ratio}. Since the ratios of tthe condensate and non-condensate charge over total charge is bound to add up to one by charge conservation, the ratio of noncondensate to total charge hence must increase with increasing backreaction. Taking these two facts together, we deduce that the depletion at the soliton core must decrease and vanish as the system approaches a homogeneous and normal state.

\bibliography{refs}

\end{document}